\documentclass[%
 reprint,
superscriptaddress,
 amsmath,amssymb,
 aps,
]{revtex4-2}

\usepackage[textsize=tiny]{todonotes}

\usepackage{graphicx}
\usepackage{dcolumn}
\usepackage{bm}
\usepackage{mathtools}

\renewcommand{\eqref}[1]{Eq.~(\ref{#1})}

\newcommand{\figref}[1]{Fig.~\ref{#1}}

\newcommand{\refcite}[1]{Ref.~\cite{#1}}

\begin{document}

\title{Competition between self-assembly and phase separation governs high-temperature condensation of a DNA liquid}

\author{Omkar Hegde}
\email{equal contribution}
\affiliation{Martin A.~Fisher School of Physics, Brandeis University, Waltham, MA 02453, USA}
\author{Tianhao Li}%
\email{equal contribution}
\affiliation{Department of Chemistry, Princeton University, Princeton, NJ 08544, USA}
\author{Anjali Sharma}
\email{equal contribution}
\affiliation{Martin A.~Fisher School of Physics, Brandeis University, Waltham, MA 02453, USA}
\author{Marco Borja}
\affiliation{Martin A.~Fisher School of Physics, Brandeis University, Waltham, MA 02453, USA}

\author{William M. Jacobs}
\email{wjacobs@princeton.edu}
\affiliation{Department of Chemistry, Princeton University, Princeton, NJ 08544, USA}

\author{W. Benjamin Rogers}
\email{wrogers@brandeis.edu}
\affiliation{Martin A.~Fisher School of Physics, Brandeis University, Waltham, MA 02453, USA}

\date{\today}
 
\begin{abstract}
  In many biopolymer solutions, attractive interactions that stabilize finite-sized clusters at low concentrations also promote phase separation at high concentrations.
  Here we study a model biopolymer system that exhibits the opposite behavior: Self-assembly of DNA oligonucleotides into finite-sized, stoichiometric clusters, known as ``DNA nanostars'', tends to inhibit phase separation of the oligonucleotides at high temperatures.
  We use microfluidics-based experiments to map the phase behavior of DNA nanostars at high concentrations of divalent cations, revealing a novel phase transition in which the oligonucleotides condense upon increasing temperature.
  We then show that a theoretical model of competition between self-assembly and phase separation quantitatively predicts changes in experimental phase diagrams arising from DNA sequence perturbations.
  Our results point to a general mechanism by which self-assembly shapes phase boundaries in complex biopolymer solutions.
\end{abstract}

\maketitle

Biopolymers can assemble into higher-order structures via a wide variety of mechanisms, including aggregation~\cite{knowles2014amyloid}, physical gelation~\cite{jain2017rna}, and phase separation~\cite{shin2017liquid}.
Given the recently recognized role of phase separation in regulating a host of biological processes \textit{in vivo}~\cite{hyman2014liquid}, it is important to understand how these mechanisms influence one another, for example through coupled gelation and phase separation of prion-like domains~\cite{li2012phase} or via phase separation-accelerated assembly of pathological fibrils~\cite{lipinski2022biomolecular,weber2019spatial}, viral capsids~\cite{guseva2020measles,hagan2023self}, and other biological structures~\cite{mcdonald2020assembly,mccall2018partitioning}.

One important scenario is the interplay between liquid--liquid phase separation (LLPS) and the self-assembly of finite-sized clusters, commonly referred to as \textit{oligomers}, in the dilute phase.
In simple fluids, the tendency for atoms and small molecules to form clusters at low concentrations is directly related to the critical temperature for condensation~\cite{mcquarrie1976statistical}.
This relationship, which is typically quantified using virial coefficients, has been shown to hold for molecular fluids~\cite{mcquarrie1976statistical}, colloidal particles~\cite{vliegenthart2000predicting}, and globular proteins~\cite{thomson1987binary,neal1998molecular} interacting via potentials that are effectively pairwise in nature.

However, in solutions of disordered polypeptides and nucleic acids with greater conformational heterogeneity, the relationship between oligomerization and phase separation may not be as simple.
Recent examples include cases in which disordered polypeptides are clustered to a greater extent in the dilute phase than anticipated on the basis of simple fluids~\cite{kar2022phase,lan2022quantitative}, as well as cases in which the formation of protein oligomers in the dilute phase tends to oppose their phase separation into ribonucleic condensates~\cite{seim2022dilute}.
Theoretical and experimental studies have also suggested that the assembly of stoichiometric oligomers can modulate phase boundaries by hiding binding interfaces that would otherwise contribute to LLPS~\cite{reinhardt2011re,rovigatti2013self,jacobs2014phase,sanders2020competing}.
However, experimental systems in which these ideas can be tested via \textit{de novo} design of biopolymers and directly compared with theoretical predictions are lacking.

\begin{figure}[t]
  \includegraphics[width=\columnwidth]{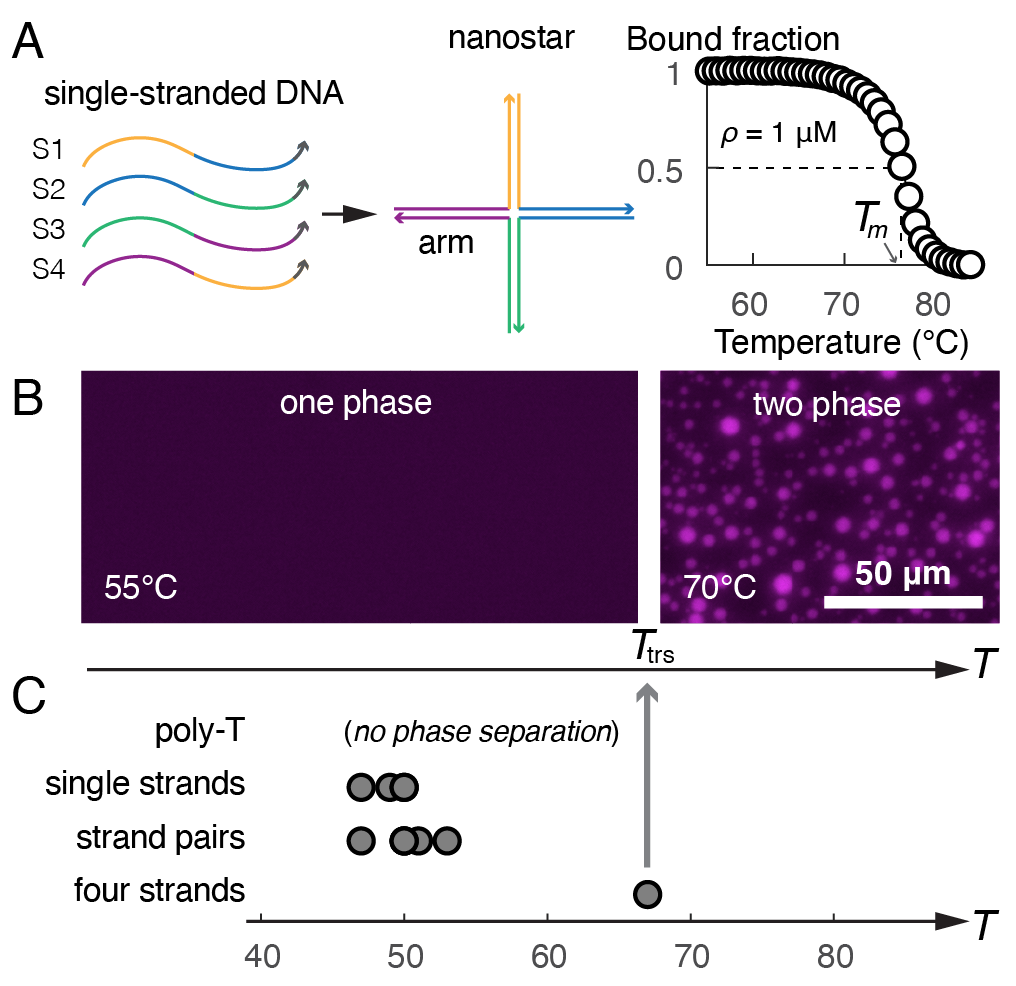}
  \caption{\textbf{Nanostar-forming DNA liquids undergo sequence-dependent condensation at high divalent counter-ion concentrations.}
    (A)~Nanostars self-assemble from a mixture of four sequences in dilute solution.  UV-absorbance measurements of the bound fraction of DNA bases at dilute DNA concentration ($\rho = 1\,\mu\text{M}$ DNA) indicate a nanostar melting temperature of approximately $T_{\text{m}} \simeq 75^\circ\text{C}$.
    (B)~At high DNA and divalent counter ion concentrations ($200\,\mu\text{M}$ DNA, $100\,\text{mM}$ Mg$^{2+}$), fluorescence microscopy reveals a high-temperature phase transition.
    (C)~Measurements of the transition temperature, $T_{\text{trs}}$, indicate that condensation is strongly sequence and stoichiometry-dependent.
    \label{fig:1}}
    
\end{figure}

To address this gap, we present a model DNA-based system to systematically explore the interplay between phase separation and oligomer self-assembly.
A crucial advantage of our experimental system is the ability to control the oligomer stability systematically via sequence design, enabling direct comparisons with the predictions of theoretical models.
We first show that DNA solutions that self-assemble into ``nanostars''~\cite{biffi2013phase,jeon2018salt}---stoichiometric oligomers with tunable free energies of formation---can undergo a novel phase transition at high MgCl$_2$ concentrations, in which the DNA solution condenses upon increasing temperature.
We then introduce a microfluidics-based method to infer the phase boundaries of the phase-separated solution.
Finally, using a combination of theory and experiment, we show that this phase behavior is a consequence of competition between nanostar self-assembly and LLPS of disassembled DNA strands.
We discuss how this mechanism is distinct from that of thermosensitive polymer solutions with lower critical solution temperatures (LCSTs).

We study a mixture of four distinct DNA sequences that self-assemble into four-arm nanostars~\cite{biffi2013phase,jeon2018salt} in a solution containing $100~\,\text{mM}$ Mg$^{2+}$ counter ions.
Each arm consists of a 20-base-pair double helix joining two strands of the four-strand nanostar (\figref{fig:1}a and Table~S1 in the Supplementary Material~\cite{SI}). 
In dilute solution, nanostars assemble upon cooling (\figref{fig:1}a).
Yet at higher DNA concentrations, we unexpectedly find that the solution phase-separates upon heating (\figref{fig:1}b), leading to the formation of coexisting dense and dilute phases that are enriched and depleted in DNA, respectively.
Condensation occurs at a transition temperature, $T_{\text{trs}} \simeq 67^\circ\text{C}$, in a solution with $200\,\mu\text{M}$ DNA, which is near to the melting temperature, $T_{\text{m}} \simeq 75^\circ\text{C}$, of the nanostars in dilute solution.
Importantly, this phase transition takes place consistently whether the sample is heated or cooled, indicating that we are measuring the equilibrium phase behavior of the solution~\cite{SI}.

To understand the determinants of this previously unreported phase transition, we vary the cations, DNA sequences, and stoichiometry of the solution.
We find that condensation requires a relatively high concentration of $\text{Mg}^{2+}$ ions ($[\text{Mg}^{2+}] > 50\,\text{mM}$~\cite{SI}), and that exchanging $\text{Mg}^{2+}$ for a monovalent $\text{Na}^+$ cation prevents phase separation completely.
However, the phase behavior cannot be explained solely by the counter ion identity and concentration.
First, poly-T sequences of the same length remain in a single phase across the entire temperature range that we probe ($T \le 75^\circ \text{C}$) (Fig.~\ref{fig:1})~\cite{SI}.
This striking sequence dependence contrasts with the sequence-independent counter ion-induced condensation of higher valent polyamines like spermine and spemidine~\cite{braunlin1982equilibrium,raspaud1998precipitation}.
The fact that $T_{\text{trs}} < T_{\text{m}}$ in our system also implies that DNA base-pairing occurs with high probability in the temperature range where condensation occurs.
These features suggest that phase separation in our system is driven by a combination of cation-dependent and base-pairing interactions.
Second, we find that $T_{\text{trs}}$ depends strongly on the stoichiometry of the solution.
Solutions comprising either single strands (S1--S4 in \figref{fig:1}a) or 50\%-50\% mixtures of strand pairs, all of which have nearly the same overall G/C content as the stoichiometric four-strand nanostar solution, phase-separate at temperatures approximately $15^\circ\text{C}$ lower than the nanostar solution $T_{\text{trs}}$ (\figref{fig:1}c).
These observations suggest that the phase boundary is influenced by the ability of the strands to self-assemble when mixed in stoichiometric proportion.

\begin{figure}[t]
  \includegraphics[width=\columnwidth]{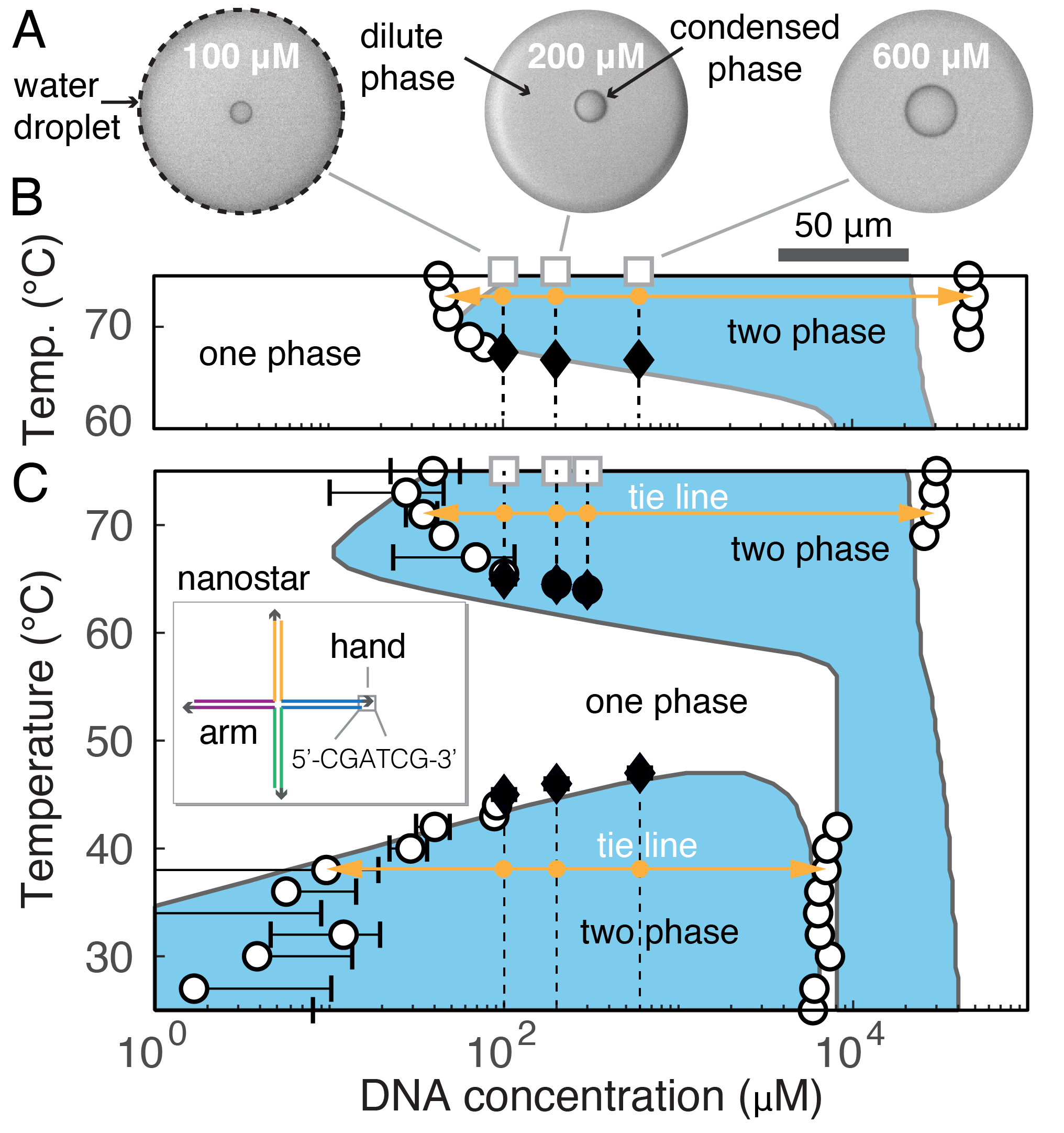}
  \caption{\textbf{Microfluidics-based experiments map the complete phase diagram.}
    (A)~Bright-field images of the phase-separated condensate within emulsion droplets of known total DNA concentration are used to infer the dilute and condensed-phase DNA concentrations. The three images were taken at $75^\circ$C.
    (B)~Comparison of theoretically predicted (solid lines and shaded coexistence regions) and experimentally measured phase diagrams (symbols) for the system shown in \figref{fig:1}a (nanostars without hand-shaking sequences).  Circles indicate the binodal inferred by applying the lever rule to condensate volume fractions measured at three total DNA concentrations (vertical dashed lines); diamonds indicate direct measurements of the transition temperatures.
    (C)~The complete phase diagram for the system depicted in the inset (nanostars with hand-shaking sequences), which has a second coexistence region at low temperatures.  Error bars indicate uncertainties in the inferred binodals, as described in~\cite{SI}.
    \label{fig:2}}
\end{figure}

We next introduce a microfluidics-based platform to determine the concentrations of coexisting dilute and condensed phases~\footnote{While this manuscript was in preparation, a similar method for computing LLPS phase boundaries via microfluidics was reported in \refcite{villois2022droplet}}.
We create monodisperse water droplets in oil, which are filled with a user-specified concentration of DNA in 100~mM MgCl$_2$/1xTE, and image them in a sealed capillary~\cite{Mcdonald2000,Hensley2022}.
After letting the droplets equilibrate at a given temperature, we measure the condensate and droplet volumes using image analysis to determine the condensate volume fraction~\cite{SI}.
At equilibrium, the condensed phase forms a single spherical droplet in an emulsion droplet of known total DNA concentration (\figref{fig:2}a).
By measuring the radii of dozens of condensed-phase droplets at three different total concentrations at the same temperature, we can infer the bulk-phase concentrations on the binodal by applying the lever rule, $x_{\text{l}} \rho_{\text{l}} + (1 - x_{\text{l}}) \rho_{\text{v}} = \rho$, where $\rho_{\text{l}}$ and $\rho_{\text{v}}$ are the unknown concentrations of the condensed and vapor phases, respectively, $x_{\text{l}}$ is the measured condensed-phase volume fraction, and $\rho$ is the total DNA concentration in the emulsion droplet.
In practice, we find that the condensed-phase volume fractions in the phase-separated regions are not linear functions of the total DNA concentration~\cite{SI}, as expected for a single-component solution.
We hypothesize that this non-linearity is a consequence of length polydispersity arising from a combination of prematurely truncated DNA strands and imperfectly assembled nanostars~\cite{biffi2013phase}.
We therefore account for the known strand-length polydispersity in our system~\cite{SI} to infer the complete binodals (circles in \figref{fig:2}b).
We also directly observe the transition temperatures at which each total concentration intersects the dilute arm of the binodal (diamonds in \figref{fig:2}b).
In this way, we can map the complete phase diagram of the four-strand mixture.

We further validate our microfluidics-based method by mapping the complete phase diagram of the well-studied variety of nanostars with self-complementary ``hand-shaking'' sequences (\figref{fig:2}c).
In this system, a 6-base palindromic hand-shaking sequence is appended to the end of each strand, resulting in a dangling end that can hybridize with another such sequence on a different assembled nanostar (inset of \figref{fig:2}c).
These hand-shaking interactions cause assembled nanostars to phase-separate below an upper critical solution temperature (UCST) of approximately 45$^\circ$C.
The resulting low-temperature phase diagram is consistent with prior studies of DNA nanostar LLPS under different salt conditions~\cite{biffi2013phase,jeon2018salt,Saleh2022}, in which UCST-like phase separation has been rationalized in terms of a ``patchy-particle'' model of assembled nanostars as rigid, colloidal particles~\cite{rovigatti2014accurate,locatelli2017condensation}.
Importantly, adding hand-shaking sequences does not measurably affect the novel high-temperature phase transition.
The fact that the low and high-temperature phase transitions are essentially orthogonal to one another suggests that the high-temperature condensed phase does not comprise assembled nanostars.
Consistent with this hypothesis, we find that the DNA concentration of the high-temperature condensed phase is approximately five times greater ($\sim 450~\text{mg/ml}$) than that of the well-studied low-temperature condensed phase of assembled nanostars ($\sim 100~\text{mg/ml}$)~\cite{rovigatti2014accurate,locatelli2017condensation} (\figref{fig:2}c), suggesting that the higher-concentration condensed phase associated with high-temperature condensation resembles a polymer melt of disassembled DNA strands~\cite{rubinstein2003polymer}.

\begin{figure}[t]
  \includegraphics[width=\columnwidth]{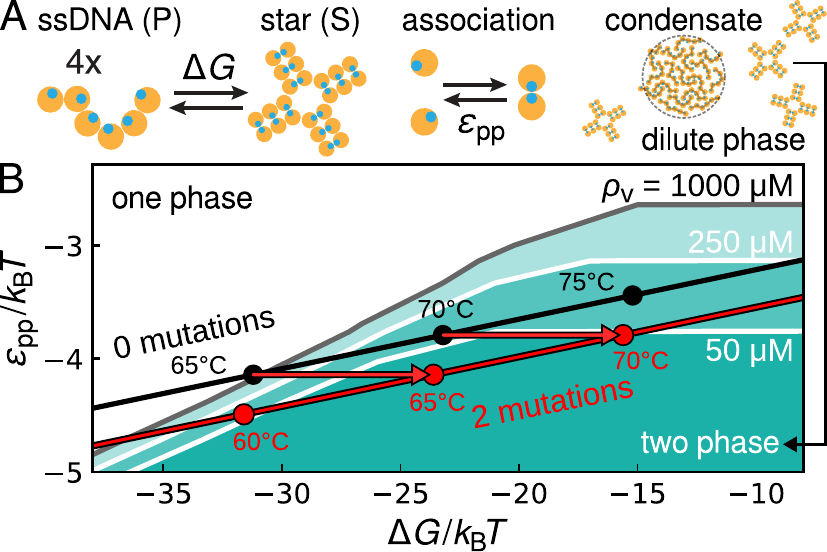}
  \caption{\textbf{A coarse-grained theoretical model captures the competition between LLPS and self-assembly.}
    (A)~Single-stranded DNA (ssDNA) is described by chains of Kuhn-segment blobs (gold circles).  Nanostar self-assembly is modeled by a two-state reaction with free-energy change $\Delta G$.  Blobs associate via one-to-one binding sites (blue circles) according to the association free energy $\epsilon_{\text{pp}}$.  A polymer melt of disassembled strands coexists with a dilute phase of assembled nanostars as $\beta\Delta G \rightarrow -\infty$.
    (B)~The high-temperature phase behavior predicted by this model is governed by $\Delta G$ and $\epsilon_{\text{pp}}$.  Shaded regions indicate predicted two-phase regions with different dilute-phase DNA concentrations, $\rho_{\text{v}}$.  Red and black lines indicate the temperature-dependence of $\Delta G$ and $\epsilon_{\text{pp}}$ for nanostars with and without two destabilizing mutations per nanostar arm, respectively (cf.~\figref{fig:4}a).  The transition temperature, $T_{\text{trs}}$, is found where the parameterization curve intersects a prescribed total DNA concentration, $\rho = \rho_{\text{v}}$.
    \label{fig:3}}
\end{figure}

Taken together, our observations suggest that the high-temperature phase transition that we observe arises from a competition between nanostar assembly and LLPS of disassembled DNA strands.
Specifically, we propose that a combination of base-pairing and divalent counter-ion mediated interactions provide the driving force for phase separation at high temperatures.
However, lowering the temperature favors the self-assembly of discrete nanostars due to strong, complementary base-pairing interactions.
Nanostar self-assembly reduces the concentration of disassembled strands that can undergo phase separation, leading to a transition from two phases to one phase at a temperature slightly below the nanostar melting temperature $T_{\text{m}}$.

To formalize this hypothesis, we introduce a mean-field theoretical model based on statistical associating fluid theory~\cite{chapman1989saft} in combination with a two-state model of nanostar formation (\figref{fig:3}a).
We treat single-stranded DNA (ssDNA) as freely jointed chains of $N_{\text{p}}$ blobs, each representing an approximately 7-nucleotide-long Kuhn-segment with excluded volume $v_0 \approx 4\,{\text{nm}}^3$~\cite{CHI20131072,MURPHY20042530}.
Each blob can associate reversibly with at most one other blob with association free energy $\epsilon_{\text{pp}}$.
Nanostar self-assembly follows a two-state model,
\begin{equation}
  \label{eq:two-state}
  n\text{P} \xrightleftharpoons{\Delta G} \text{S},
\end{equation}
where P represents a disassembled strand, S indicates an assembled nanostar, $n=4$ accounts for the nanostar stoichiometry, and $\Delta G$ is the free energy of nanostar formation.
\eqref{eq:two-state} implies that ssDNA exclusively exists in either a disassembled state or as part of a fully assembled nanostar.
When in the S state, ssDNA is considered to be fully associated and is thus not allowed to engage in further interactions.
Hand-shaking sequence blobs, which are included to allow us to model the UCST of assembled nanostars (\figref{fig:2}c), are the exception to this rule when they are present in the system; see \cite{SI}.
With these ingredients, we write down a mean-field free-energy density and calculate phase coexistence by satisfying $ n\text{P}\rightleftharpoons\text{S}$ chemical equilibrium as well as equal chemical potentials and pressures among coexisting phases~\cite{li2023interplay}.
As anticipated, this model predicts that the high-temperature phase boundary is primarily determined by the parameters $\Delta G$ and $\epsilon_{\text{pp}}$ in nanostar solutions both with and without hand-shaking sequences.
We therefore summarize the predicted high-temperature phase behavior in \figref{fig:3}b, which shows whether a solution with a prescribed total DNA concentration forms one or two phases as a function of $\beta\Delta G$ and $\beta\epsilon_{\text{pp}}$.
(See \cite{li2023interplay} for a detailed description of this modeling approach).

We next determine how the interaction parameters depend on temperature in a particular DNA system.
Since these parameters derive from the same base-pairing and counter-ion-dependent interactions, we assume that $\epsilon_{\text{pp}}(T)$ and $\Delta G(T)$ are linearly related, such that $\Delta G$ represents a sum of associative interactions and a temperature-independent entropic penalty for nanostar formation.
Moreover, both of these free energies are expected to be approximately linear functions of temperature~\cite{santalucia1998unified}.
Within these physically motivated assumptions, we find that our model can reproduce the high-temperature binodal (\figref{fig:2}b,c) while also predicting satisfactory agreement with the nanostar $T_{\text{m}}$ in dilute solution.
The resulting temperature-dependence of $\beta\epsilon_{\text{pp}}(T)$ and $\beta\Delta G(T)$ (black curve in \figref{fig:3}b) is also compatible with the hybridization free-energy predictions of NUPACK~\cite{NUPACK2020}.
Finally, we determine the hand-shaking interaction parameter $\epsilon_{\text{ss}}(T)$ from NUPACK to predict the UCST in systems with hand-shaking sequences (\figref{fig:2}c).
(See \cite{SI} for a detailed analysis of our parameterization approach.)

\begin{figure}[t]
  \includegraphics[width=\columnwidth]{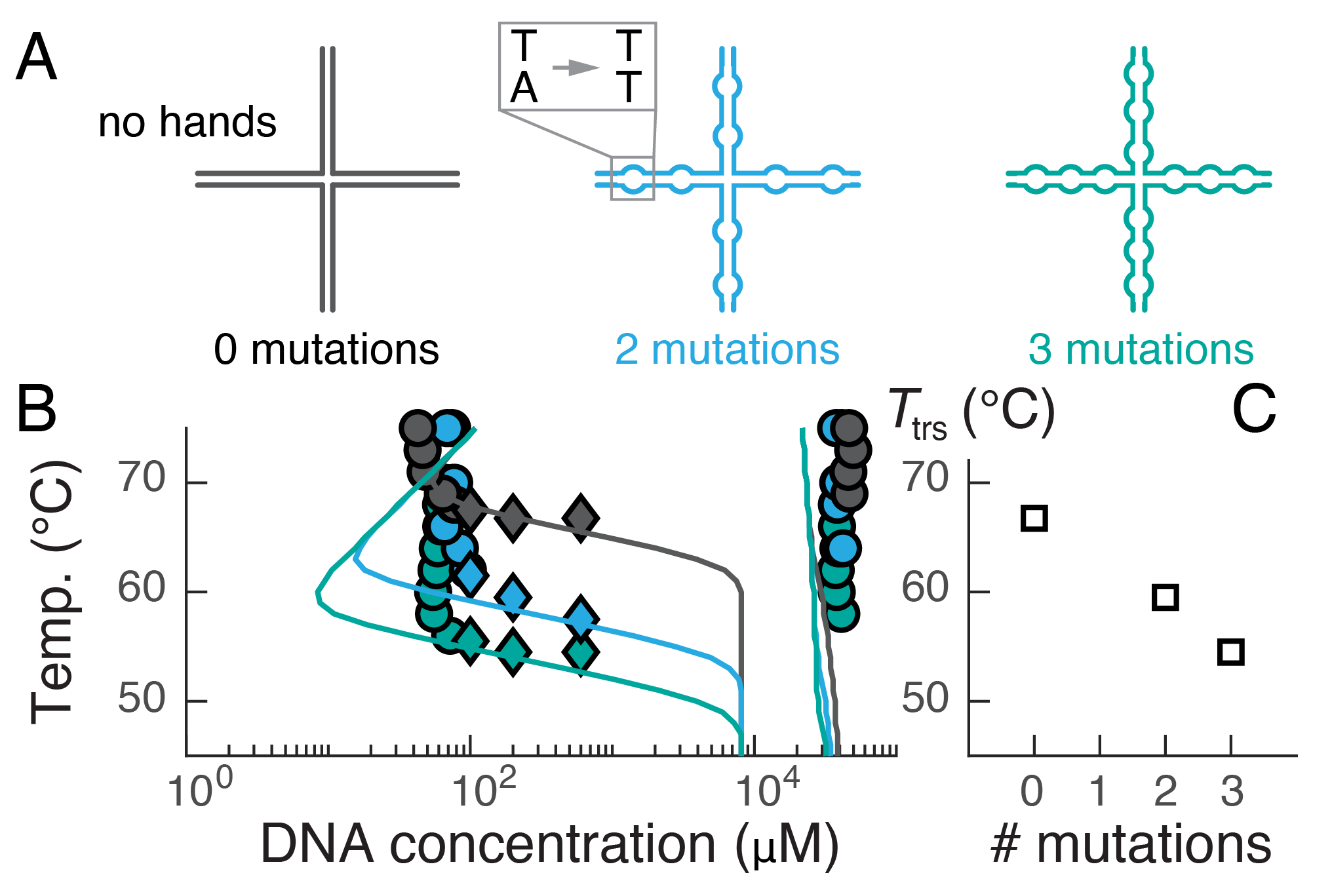}
  \caption{\textbf{Tuning the stability of individual nanostars shifts the balance between self-assembly and LLPS.}
    (A)~Introducing bulges into nanostars without hand-shaking interactions shifts $\Delta G$ while negligibly affecting $\epsilon_{\text{pp}}$.
    (B)~Comparison of theoretical (solid lines) and experimental (symbols) phase diagrams for systems with 0, 2, and 3 mutations per arm.  Circles and diamonds indicate inferred and direct binodal measurements, respectively, as in \figref{fig:2}b,c.
    (C)~The experimentally determined transition temperatures at $200\,\mu\text{M}$ DNA depend linearly on the number of mutations per arm, $m$, as predicted.
    \label{fig:4}}
\end{figure}

To test the predictive capabilities of our model, we exploit our ability to tune the nanostar stability via sequence design.
We systematically destabilize nanostars (without hand-shaking sequences) by introducing point mutations into the arms, which lead to ``bulges''~\cite{santalucia1998unified} when assembled (\figref{fig:4}a).
$\Delta G(T)$ increases linearly with the number of mutations per arm, $m$,
\begin{equation}
  \label{eq:mutations}
  \Delta G(T;m) \approx \Delta G(T;0) + n m \Delta\Delta G,
\end{equation}
where the free-energy penalty per mutation is $\Delta\Delta G \approx 0.95~k_{\text{B}}T$~\cite{SI}.
Importantly, these mutations do not affect the G/C content of the sequences, and thus have a negligible effect on the associative interactions, $\epsilon_{\text{pp}}$, between ssDNA blobs.
As a result, the temperature-parameterization curve, $(\beta\Delta G(T), \beta\epsilon_{\text{pp}}(T))$, of mutated nanostars is shifted to the right (red arrows) in \figref{fig:3}b.
We therefore predict that the transition temperatures for high-temperature phase separation will be systematically shifted as well.

Mapping the phase diagrams of these nanostar variants using our microfluidics-based platform reveals good agreement with the predictions of our theoretical model (\figref{fig:4}b).
We obtain additional theoretical insight by considering the relevant limit of stable nanostars ($\beta\Delta G \rightarrow -\infty$; \figref{fig:3}a), which leads to an approximate linear relation for the phase boundary~\cite{SI},
\begin{equation}
  \label{eq:phase-boundary}
  \beta\Delta G - \ln\left(\frac{n}{\rho_{\text{v}} v_{\text{s}}}\right) = \frac{n N_{\text{p}} \beta\epsilon_{\text{pp}}}{2} - \ln \left[ \frac{(1 - \phi_{\text{l}})^{nN_{\text{p}}}}{\phi_{\text{l}}^{n(1-N_{\text{p}}/2)}} \right],
\end{equation}
where $\beta \equiv 1/k_{\text{B}}T$ is the inverse temperature, $v_s$ is the excluded volume of an assembled nanostar, and the polymer volume fraction in the condensed phase, $\phi_{\text{l}} \equiv \rho_{\text{l}} N_{\text{p}} v_0$, is a constant determined by $N_{\text{p}}$.
Together with \eqref{eq:mutations} and the assumed linear dependence of $\Delta G$ and $\epsilon_{\text{pp}}$ on temperature, \eqref{eq:phase-boundary} predicts that $T_{\text{trs}}$ decreases linearly with the number of mutations, $m$.
This nontrivial prediction of our theory is validated by the experimentally measured transition temperatures (\figref{fig:4}c), which vary by $\sim 15^\circ\text{C}$ despite the minimal changes to the DNA sequences.
By contrast, varying the strand lengths by a factor of two has a relatively small effect on the transition temperature, in line with the predictions of our model~\cite{SI}.

In summary, we have reported a novel phase transition in which a DNA solution phase-separates upon heating.
Using a combination of experiments and theoretical modeling, we have demonstrated that we can control the phase boundary by tuning the stability of individual DNA nanostars, whose self-assembly opposes phase separation.
Importantly, we have shown that this mechanism is extremely sensitive to the DNA sequences and stoichiometry.
This sensitivity distinguishes our proposed mechanism from canonical lower critical solution temperature (LCST) behavior, in which effective attractive interactions among polymers that arise from cation and solvent entropic effects drive condensation at high temperatures~\cite{rubinstein2003polymer}.
While canonical LCST behavior can account for the temperature and cation-dependence of the phase boundary, it does not predict the systematic dependence of the phase boundary on G/C content-preserving point mutations (\figref{fig:4}).
A second distinguishing feature of our model is that it predicts a shift in the composition of the dilute phase from mostly ssDNA to mostly assembled nanostars as the temperature is lowered~\cite{li2023interplay}; this prediction could be tested in future experiments.

In fact, our model~\cite{li2023interplay} and related works~\cite{reinhardt2011re,bartolucci2021controlling} do not predict the existence of an LCST \textit{at all} when self-assembly competes with LLPS.
More specifically, our model predicts that the high-temperature coexistence region extends to low temperatures at very high DNA concentrations.
This feature appears because the excluded volume assigned to the nanostar is greater than that of its constituent ssDNA polymers, which is necessary to reproduce the disparate concentrations of the low-temperature (assembled nanostar-dominated) and high-temperature (disassembled strand-dominated) condensed phases when hand-shaking sequences are present (\figref{fig:2}c).
While we are unable to investigate this region of the phase diagram experimentally, and it is possible that the two-state approximation, \eqref{eq:two-state}, is insufficient under these conditions, this uncertainty does not affect our central results.

We expect that the mechanism that we describe here may apply more broadly in more complex systems where the formation of stable oligomers effectively conceals the attractive interactions that drive LLPS.
Given numerous examples of oligomer-forming protein and RNA species identified in intracellular ribonucleic condensates~\cite{sanders2020competing,guillen2020rna}, it is conceivable that this mechanism plays a role in modulating phase behavior \textit{in vivo}.
We note that the ability of divalent cations to condense nucleic acids at high temperatures, with unexpectedly strong sequence-dependent effects, has been observed in RNA solutions as well~\cite{Wadsworth2022.10.17.512593}, suggesting that similar mechanisms may be at play.
Our proposed mechanism is also related to the competition between intra- and intermolecular base-pairing that gives rise to re-entrant phase behavior in RNA solutions~\cite{kimchi2023uncovering}, whereby strong intramolecular base-pairing reduces the concentration of strands that are available for phase separation.
Overall, we anticipate that the microfluidics platform and coarse-grained modeling approach described here might generate useful insights across this broad class of biopolymer systems.

\bibliography{references.bib}

\begin{acknowledgments}
We acknowledge Daniel Hariyanto and Melissa Rinaldin for first observing the high-temperature phase transition. We also acknowledge Mohammad Nosherwon Malik for his help with the bulk fluorescence microscopy experiments. We thank Michael Hagan, Rees Garmann, Thomas Videb{\ae}k, and Fan Chen for their comments on the manuscript. This work was supported in part by grants from the Human Frontier Science Program (RGP0029) and the Smith Family Foundation to WBR, as well as support from the National Science Foundation (DMR-2143670) to WMJ.
\end{acknowledgments}

\end{document}


\title{Supplemental Information for ``Competition between self-assembly and phase separation governs high-temperature condensation of a DNA liquid''}

\maketitle

\makeatletter 
\renewcommand{\thefigure}{S\arabic{figure}}
\renewcommand{\thetable}{S\arabic{table}}
\renewcommand{\theequation}{S\arabic{equation}}

\section{Experimental methods and analysis}

\subsection{DNA sequences}

\begin{table}[h!]
  \begin{center} \vskip-3ex
    \begin{tabular}{c c} 
      \hline
      Seq.No. & Sequence\\ 
      \hline
      1 & 5’- GGACCGCGTCTCTATGGAGCAACGGCATTAGACTGCTTAGCCA-3’\\
      \hline
      2 & 5’- GCGTAGTGGCTGGCCTCGGCAAGCTCCATAGAGACGCGGTCCA-3’\\
      \hline
      3 & 5’- CCGCCAGGAACTTCGCTCGGAAGCCGAGGCCAGCCACTACGCA-3’\\
      \hline 
      4 & 5’-GGCTAAGCAGTCTAATGCCGAACCGAGCGAAGTTCCTGGCGGA-3’
    \end{tabular}
  \end{center} \vskip-3ex
  \caption{DNA sequences without sticky ends.}
\end{table}


\begin{table}[h!]
  \begin{center} \vskip-3ex
    \begin{tabular}{c c}
      \hline
      Seq.No. & Sequence\\ 
      \hline
      1 & 5’-GGACCGCGTCTCTATGGAGCAACGGCATTAGACTGCTTAGCCA\textbf{CGATCG}-3’\\
      \hline
      2 & 5’-GCGTAGTGGCTGGCCTCGGCAAGCTCCATAGAGACGCGGTCCA\textbf{CGATCG}-3’\\
      \hline
      3 & 5’- CCGCCAGGAACTTCGCTCGGAAGCCGAGGCCAGCCACTACGCA\textbf{CGATCG}-3’\\
      \hline
      4 & 5’- GGCTAAGCAGTCTAATGCCGAACCGAGCGAAGTTCCTGGCGGA\textbf{CGATCG}-3’
    \end{tabular}
  \end{center} \vskip-3ex
  \caption{DNA sequences with sticky ends (highlighted in bold text).}
\end{table}


\begin{table}[h!]
  \begin{center} \vskip-3ex
    \begin{tabular}{c c} 
      \hline
      Seq.No. & Sequence\\ 
      \hline
      1 & 5’-GGACCGCGTCTCTATGGAGCAACGGC\textbf{T}TTAGACTGC\textbf{A}TAGCCA-3\\
      \hline
      2 & 5’-GCGTAGTGGCTGGCCTCGGCAAGCTCC\textbf{T}TAGAG\textbf{T}CGCGGTCCA-3’\\
      \hline
      3 & 5’-CCGCCAGGAACTTCGCTCGGAAGCCG\textbf{T}GGCCAGCCAC\textbf{A}ACGCA-3’\\
      \hline
      4 & 5’-GGCTAAGCAGTCTAATGCCGAACCG\textbf{T}GCGAAG\textbf{A}TCCTGGCGGA-3’
    \end{tabular}
  \end{center} \vskip-3ex
  \caption{DNA sequences with two mutations (highlighted in bold text).}
\end{table}

\begin{table}[h!]
  \begin{center} \vskip-3ex
    \begin{tabular}{c c} 
      \hline
      Seq.No. & Sequence\\ 
      \hline
      1 & 5'-GGACCGCGTCTCTATGGAGCAACGGC\textbf{T}TTAG\textbf{T}CTGC\textbf{A}TAGCCA-3'\\
      \hline
      2 & 5'-GCGTAGTGGCTGGCCTCGGCAAGCTCC\textbf{T}TAGAG\textbf{T}CGCGG\textbf{A}CCA-3'\\
      \hline
      3 & 5'-CCGCCAGGAACTTCGCTCGGAAGCCG\textbf{T}GGCC\textbf{T}GCCAC\textbf{A}ACGCA-3'\\
      \hline
      4 & 5'-GGCTAAGCAGTCTAATGCCGAACCG\textbf{T}GCGAAG\textbf{A}TCC\textbf{A}GGCGGA-3'
    \end{tabular}
  \end{center} \vskip-3ex
  \caption{DNA sequences with three mutations (highlighted in bold text).}
\end{table}

\begin{table}[h!]
  \begin{center} \vskip-3ex
    \begin{tabular}{c c} 
      \hline
      Seq.No. & Sequence\\ 
      \hline
      1 & 5'- CTTTGCGACGAAGCGATAAGCGA\textbf{CGATCG}-3'\\
      \hline
      2 & 5'-CGAGCGGCTCAACGTCGCAAAGA\textbf{CGATCG}-3'\\
      \hline
      3 & 5'-CGCTTATCGCAAGCCTTCGGCGA\textbf{CGATCG}-3'\\
      \hline
      4 & 5'-CGCCGAAGGCAAGAGCCGCTCGA\textbf{CGATCG}-3'
    \end{tabular}
  \end{center} \vskip-3ex
  \caption{10 base pair arm nanostars with sticky ends (highlighted in bold text) .}
\end{table}

\begin{table}[h!]
  \begin{center} \vskip-3ex
    \begin{tabular}{c c} 
      \hline
      Seq.No. & Sequence\\ 
      \hline
      1 & 5'-GCGTGCCGAGTGCGCAACGGAGACGCGTCGCCA\textbf{CGATCG}-3'\\
      \hline
      2 & 5'-GCGAGGCCACGACGGAAGCGCACTCGGCACGCA\textbf{CGATCG}-3'\\
      \hline
      3 & 5'-GGCGACGCGTCTCCGAACGTACATGTATTGCGA\textbf{CGATCG}-3'\\
      \hline
      4 & 5'-CGCAATACATGTACGAACCGTCGTGGCCTCGCA\textbf{CGATCG}-3'
    \end{tabular}
  \end{center} \vskip-3ex
  \caption{15 base pair arm nanostars with sticky ends (highlighted in bold text).}
\end{table}

\begin{table}[h!]
  \begin{center} \vskip-3ex
    \begin{tabular}{c c} 
      \hline
      Seq.No. & Sequence\\ 
      \hline

      1 & 5'- GCACCATGTATTACACTCGGTGCACAAGTCTCACCTACAATCGATCCAAAGCA\textbf{CGATCG}-3'\\
      \hline
      2 & 5'-CGTCAGTGACTACCACGACTGTCAGAAGTGCACCGAGTGTAATACATGGTGCA\textbf{CGATCG}-3'\\
      \hline
      3 & 5'-GCTTTGGATCGATTGTAGGTGAGACAAGGATGATACGAAACCGCCAATGAACA\textbf{CGATCG}-3'\\
      \hline
      4 & 5'-GTTCATTGGCGGTTTCGTATCATCCAACTGACAGTCGTGGTAGTCACTGACGA\textbf{CGATCG}-3'
    \end{tabular}
  \end{center} \vskip-3ex
  \caption{25 base pair arm nanostars with sticky ends (highlighted in bold text).}
\end{table}

\subsection{The conditions of phase separation}
\begin{figure}[h]
\begin{center}
\includegraphics[width=0.9\columnwidth]{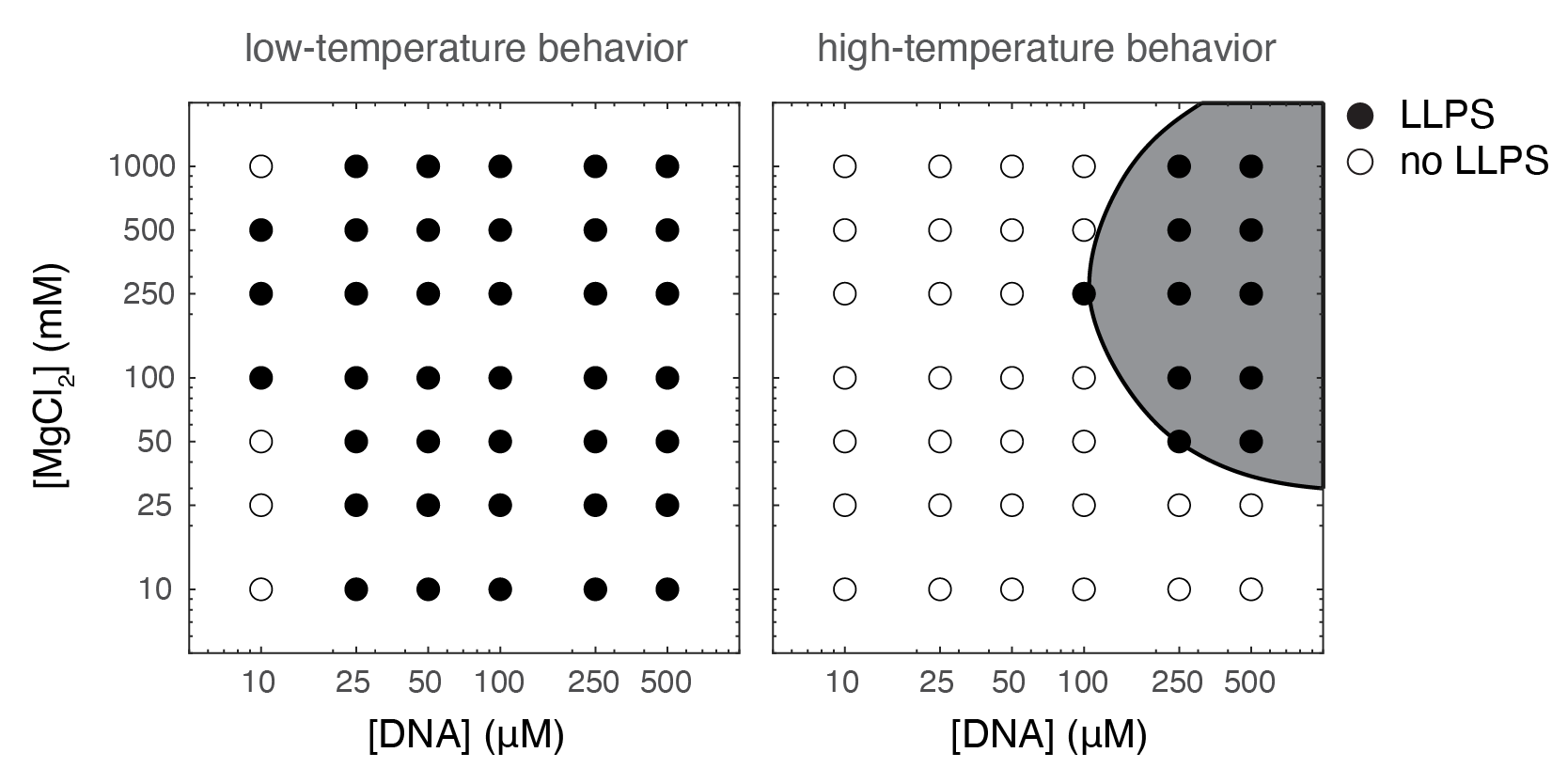}
\caption{An overview of the phase behavior of the 4-arm nanostar with sticky ends as a function of the concentration of magnesium chloride and the total concentration of DNA at both low (left) and high temperatures (right). Black points indicate that the system phase separated during a temperature sweep between 20--75 Celsius. White points indicate that the system did not phase separate.}
\label{LLPSconditions}
\end{center}
\end{figure}

We determine the conditions under which the system phase separates by performing fluorescence microscopy experiments of various DNA nanostar solutions and determining whether or not they phase separate at low temperatures, high temperatures, or both. Figure~\ref{LLPSconditions} summarizes the results of these experiments for different combinations of the magnesium concentration and the total DNA concentration. While we observe low-temperature phase separation across the full range of magnesium concentrations that we explore, we only see high-temperature phase separation at magnesium concentrations exceeding 50~mM. 

\subsection{Microfluidics platform}
Our droplet-based microfluidics experiments are accomplished in two steps: 1) fabricating the microfluidic devices; and 2) making monodisperse DNA-encapsulated droplets using the microfluidic device and measuring the size of the phase-separated condensate at different temperatures inside the droplets.

\subsubsection{Fabricating the device}
\begin{figure}[h]
\begin{center}
\includegraphics[width=0.75\columnwidth]{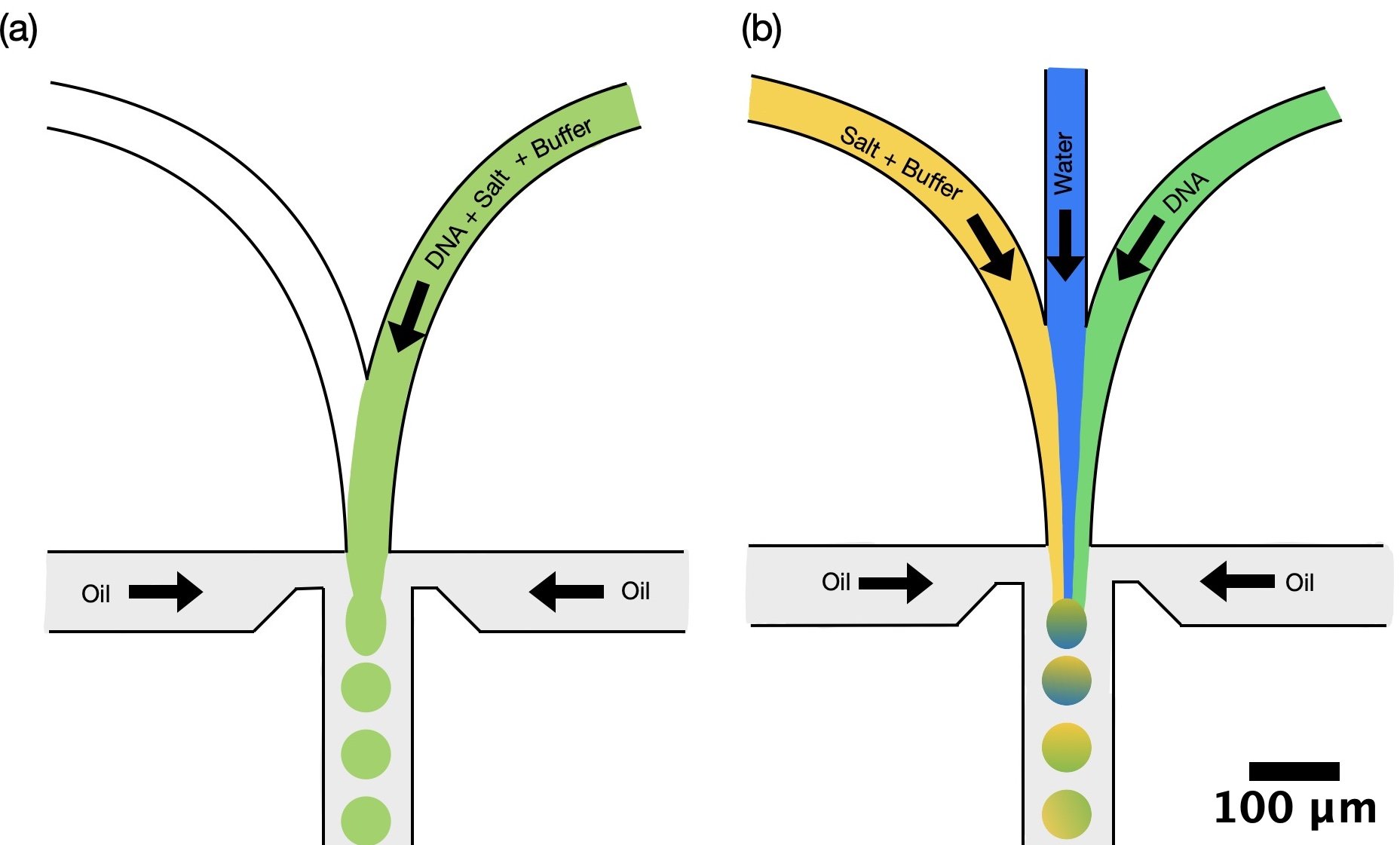}
\caption{Schematics of two microfluidic devices. Schematic (a) shows a two-channel device and schematic (b) shows a three-channel device that we use for generating droplets encapsulated with nanostars without and with hands respectively. In the three-channel device, we flow salt with buffer solution and DNA solution from the right and left channels, respectively. To prevent the mixing of DNA with a salt-buffer solution in the three-channel device, MilliQ water is flown through the middle channel.}
\label{devices}
\end{center}
\end{figure}

We use standard UV photolithographic techniques to make microfluidic devices \cite{Mcdonald2000}. Two different devices are used to obtain the data presented in the main text: 1) a two-channel device for measuring the binodal curve of the nanostars without hands; and 2) a three-channel device for measuring the binodal curve of the nanostars with hands. Schematics of the two devices are shown in \figref{devices}.

In brief, 3~ml of SU8 (SU-8 2075, MicroChem) is gently placed  on a silicon wafer (3-76-024-V-B, Silicon Materials Inc.) and spun at 500 rpm with a spin coater at a ramp rate of 100 rpm/s for 5 seconds. The spin rate is increased to 3000 rpm at a ramp rate of 300 rpm/s for 60 seconds. The film is then soft baked by placing the wafer on a hot plate at 65$^\circ$C for 3 minutes and 95$^\circ$C for 5 minutes. The above protocol ensures a uniform thickness of the film of SU-8 (with a thickness of 40~$\mu$ m) on the silicon wafer. The photo mask (which has a microfluidic pattern) is then aligned on the silicon wafer and is exposed to UV light for 50 seconds. After UV curing, the photo mask is removed and the silicon wafer is hard baked on a hot plate at 65$^\circ$C for 3 minutes and at 95$^\circ$C for 20 min.  The undeveloped photoresist is etched by dipping the wafer in a glass petri dish half-filled with propylene glycol methyl ether for around 10 minutes under continuous shaking. Once all the unreacted photoresist is dissolved in the propylene glycol methyl ether, the wafer is washed with isopropanol and dried using an air gun. 

The silicon wafer with the microfluidic device pattern on it is referred to as the master. A master is the negative of the actual device and can be used multiple times to create polydimethylsiloxane (PDMS) based microfluidic devices.
To make a PDMS device from the master, PDMS monomer (1673921, Dow Chemical Company) and cross-linker (1673921, Dow Chemical Company) are mixed in the ratio 10:1 in a plastic cup. Thorough mixing is achieved by placing the plastic cup in Thinky AR-250 Planetary centrifugal mixer for 6 minutes. 
The master with the pattern is placed in a plastic Petri dish and the monomer-cross-linker mixture is poured onto the master. To remove all of the entrapped air bubbles and gas in the mixture, the petri dish is desiccated using a vacuum desiccator. After removing the air bubbles, we place the petri dish in an oven at 70$^\circ$C for a period of three hours. The crosslinked PDMS in the oven is peeled from the master and excess PDMS is cut using a blade. Holes are punched onto the PDMS replica using a coring device 
(69039-07, Electron Microscopy Sciences) to provide for the inlet and outlet tubing in the device. The glass slide and the PDMS etched surface are subjected to oxygen plasma surface treatments (Zepto, Diener Electronic) and are bonded together to make the microfluidic device.

\subsubsection{Making the droplets}

The device channels are made hydrophobic by flushing them with Aquapel (B004NFW5EC, Amazon) and are subsequently flushed with compressed air. The entire device is then filled with HFE-7500 oil (Ran Biotechnologies). The flow rates of the aqueous and oil phases are controlled  using syringe pumps (98-2662, Harvard Apparatus). Tubes of size 0.012" inner diameter x 0.030" outer diameter (06417-11, Cole-Palmer) are used for inlets and outlets of the microfluidic device. The tube dimensions are chosen such that the tubes fit both the syringe needle (the inner diameter of the tube fits the outer diameter of the syringe needle) and the punched holes on the device (the outer diameter of the tube fits the inlets and outlets of the device). 

For the no-hands DNA, a single inlet device is sufficient, as the mixture does not phase separate at room temperature. We flow HFE-7500 with 2.5\% RAN fluorosurfactant (008-FluoroSurfactant-5wtH-20G, RAN Biotechnologies) into the oil inlet. In one of the aqueous inlets, we flow the desired concentration of DNA in 100~mM MgCl$_2$/1XTE (refer to \figref{devices}a). The other inlet is blocked using a paper clip. To create droplets with a diameter of orders of 100~$\mu$m or less, we set the aqueous flow rate to 200~$\mu$l/hour and the oil flow rate to 100~$\mu$l/hour. 

For the nanostars with hands that phase separate at room temperature, we use a device with three aqueous inlets and we flow MilliQ water in the middle inlet to prevent the mixing of the DNA with the salt and buffer. Through the other two inlets we flow DNA suspended in water through one inlet and salt buffer through the other (see \figref{devices}b). Low flow rates of water through the middle inlet lead to the mixing of DNA with salt and buffer before the formation of the droplets, leading to the clogging of the channel by the dense phase. By trial and error, we identify the optimum flow rates of each channel. For the experiments conducted here, the following flow rates are used: 90, 60, 150~$\mu$l/hour for the inlets of DNA, salt+Buffer, and water, respectively. The total volume flow rate through the three inlets is 300~$\mu$l/hour. The flow rate of oil is set to 400 ~$\mu$l/hour. 

Since the flow into the device is equal to the flow out of the device, the constituents of the droplets are mixed in proportion to the inlet flow rates. Monodisperse droplets with uniform condensate sizes are generated with this device (see \secref{sec:measuring-radii}). Therefore, we assume that the constituents of the inlets are equally distributed within the water droplets. Thus the initial DNA concentration through the inlet channel decides the final concentration of DNA contained in droplets. For example, if the inlet concentration of the DNA is 1~mM, the final concentration of DNA in all the droplets that come out of the device will be 0.3~mM. Thus, by using initial concentrations of DNA of 0.333 mM, 0.666 mM, 1 mM, and 2 mM  we get droplets containing DNA concentrations of 0.1 mM, 0.2 mM, 0.3 mM, and 0.6 mM respectively. We also confirm that the DNA concentration in each droplet is the same by verifying that the volume percent of the DNA condensate that forms within each droplet is the same. See \figref{fig:bright-field-images} for an example. Similarly, the inlet concentration of salt and buffer is adjusted such that for the given flow rate of 60 ~$\mu$l/hour (combined salt and buffer) the final droplets contain 100~mM MgCl$_2$/1XTE.  

\subsection{Preparing microscopy sample chambers}

To observe the droplets using optical microscopy, we load the droplets into a rectangular capillary with a cross-section of 0.2 X 2mm (CM Scientific). We choose a capillary height that is close to the droplet diameter to ensure that the droplets form a monolayer but do not get squeezed  by the capillary walls. For our droplets, we load 3 $\mu$l  of our emulsion using a pipette  into a 3-cm-long rectangular capillary. We cut the tip of the pipette to make the opening wider so as not to break the droplets. We fill the rest of the capillary with HFE-7500 and 2.5\% RAN fluorosurfactant, then place the capillary on a rectangular glass cover slip. After sealing both openings of the capillary with UV-curable epoxy (NOA68; Norland), we cure it under a UV lamp (UV gel nail polish dryer; Melodysusie) for 30 minutes. While keeping the sample under UV light, we cover the capillary using aluminum foil to prevent any UV damage to the DNA.  

Once the UV glue is cured, we place the slide on a rectangular acrylic frame with the capillary side facing down. This method helps to avoid any tilt in the capillary caused by the cured epoxy. To uniformly control the sample temperature, we attach a Peltier unit to the flat side of the coverslip using optical gel. A detailed description of the heater is provided in Reference \cite{Hensley2022}. 
At high temperatures, we observe that the  images of the droplets blur in roughly 10 minutes due to condensation on the inner capillary surface. In order to prevent this condensation, we surface treat the capillary by flowing aquapel through the capillary and blow dry it with a jet of air. The procedure for filling the capillaries with the droplets remains the same.

\subsection{Measuring the binodal curve}

\subsubsection{Performing the experiment}

\begin{figure}
\begin{center}
\includegraphics[width=0.75\columnwidth]{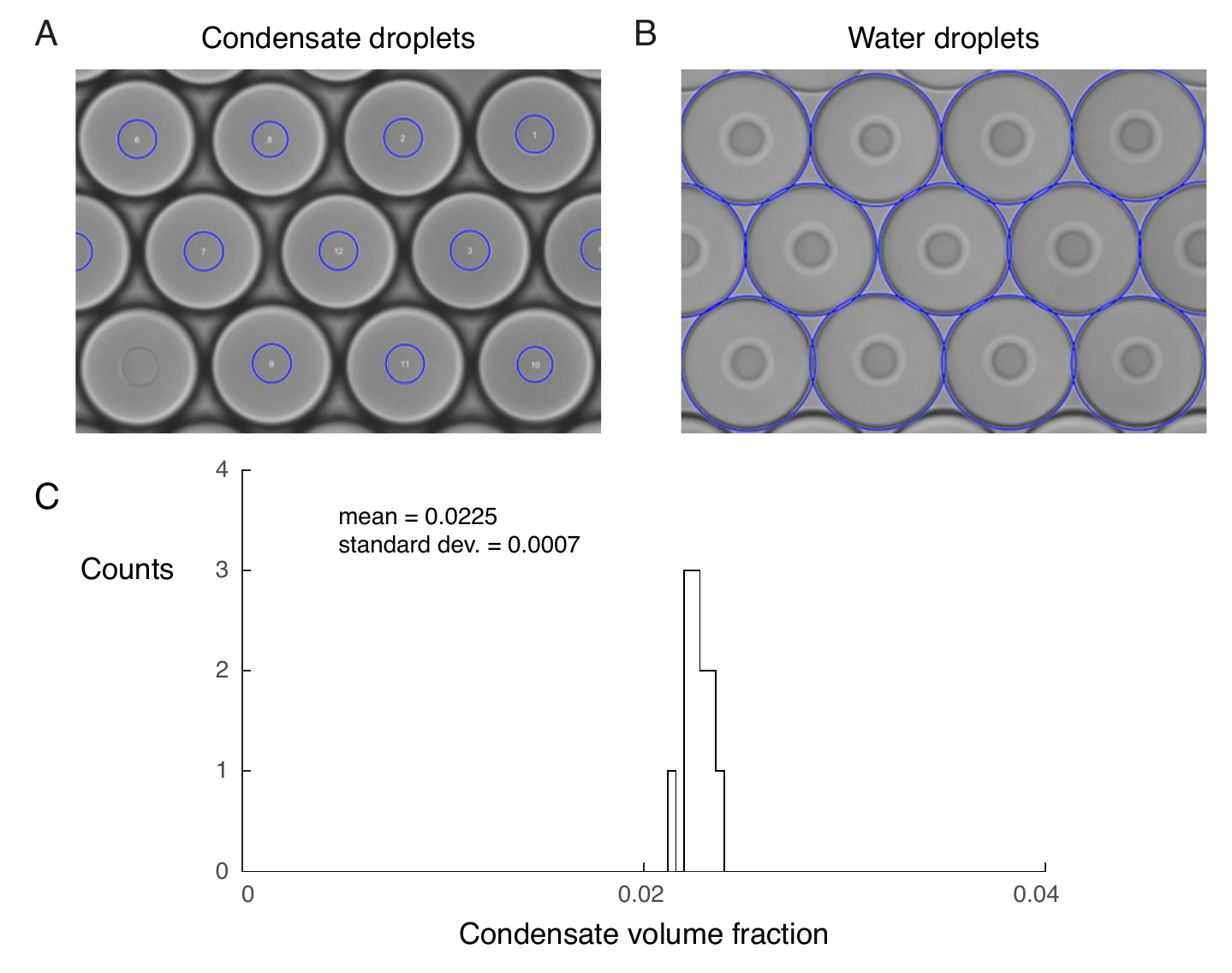}
\caption{Bright-field images of the (a) inner (liquid-phase DNA condensate) and (b) outer droplets (entire droplet). The droplets are identified (marked in blue) and their radii are measured using a MATLAB code. (c) The histogram of the volume fraction of the DNA condensate measured from images in panels a and b shows that the condensate volume fraction is nearly constant from drop to drop; the relative standard deviation is less than three percent.}
\label{fig:bright-field-images}
\end{center}
\end{figure}

\begin{figure}
\begin{center}
\includegraphics[width=0.75\columnwidth]{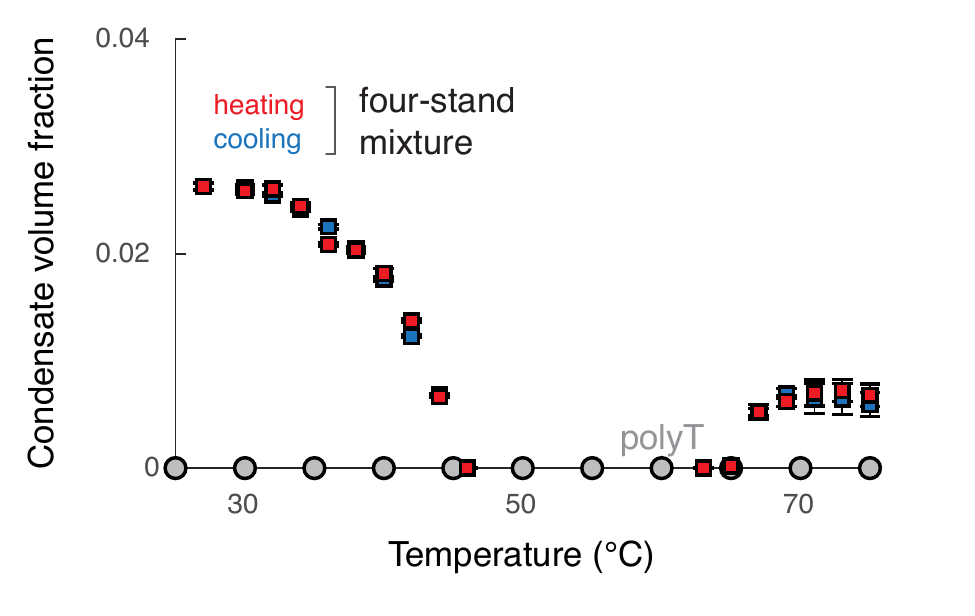} \vskip-2ex
\caption{Measured condensate volume fractions represent the equilibrium phase behavior.  Blue and red squares show the condensate volume fraction of the with-hands DNA nanostar suspension upon cooling and heating, respectively.  A DNA sequence of 50 thymines (gray circles) does not phase separate at any temperature.  All DNA concentrations are 200~$\mu$M. }
  \label{fig:heating-cooling}
\end{center}
\end{figure}

Once the droplet-filled  capillary is placed on the Peltier unit, we wait until the droplets form a monolayer and become stationary. We image the droplets using an inverted optical microscope (TE-2000; Nikon) equipped with a 10x objective and a FLIR sCMOS camera (BFLY-U3-23S6M-C; Teledyne). Next, we increase the temperature from 25 \degree C with a rate of 2$^\circ$C/min. At every 2$^\circ$C increment, we wait 5-7 minutes for the system to come to equilibrium and then record the images (\figref{fig:bright-field-images}).
To study the re-entrant phase transitions at high temperatures, we heat the sample to 75\degree C. During this event, the condensates (which are heavier) that nucleate from random sites in the droplets are allowed to settle for 10-15 minutes. Once all of the condensate nuclei fuse into one large condensate droplet, we assume that the system is in equilibrium. Then we cool the sample from 75\degree C at the rate of 2 \degree C/min until the condensate completely disappears. At every 2$^\circ$C increment, we wait 5-7 minutes for the system to come to equilibrium and then record an image. We follow the same protocol for the stars without hands. We execute the same procedure for all the temperatures until we reach the one-phase region.  We demonstrate that the system is indeed equilibrated at each temperature by performing the same experiment upon heating and upon cooling, using the same temperature-ramp protocol, and verifying that the two sets of measurements are indistinguishable, as shown in \figref{fig:heating-cooling}.

\subsubsection{Measuring the droplet radii}
\label{sec:measuring-radii}

We determine the inner and outer droplet radii using image analysis routines written in MATLAB. Since the focal planes for the outer and inner droplets are different, we first focus on the equator of the water droplets and take an image; then we focus on the equator of the inner condensate droplets and take an image. See \figref{fig:bright-field-images}a,b for examples. Images are then subjected to threshold and sharpening to increase the contrast along the edges. These steps help us identify foreground pixels of the high gradient using a Circular Hough Transform (CHT) for finding circles in images. This approach is used because of its robustness in detecting circular objects in the presence of noise, occlusion, and varying illumination. The method is employed using the imfindcircles() function in MATLAB computed using the phase-coding method first employed by T.J.~Atherton \cite{Atherton1999}. \figref{fig:bright-field-images} shows an example image with the outer and inner droplets overlaid with circles of their measured radii. 

\subsubsection{Inferring the coexisting phase concentrations}
In order to determine the concentrations of the coexisting condensed (liquid, L) and dilute (gas, G) phases, we make use of the lever rule,
\begin{equation}
  \label{eq:lever}
  \rho^{\text{L}} x + \rho^{\text{G}} (1 - x) = \rho,
\end{equation}
where $\rho$ is the total DNA concentration and $x$ is the volume fraction of the emulsion droplet occupied by the condensed phase.
To this end, we measure the radii of the inner phase-separated droplets at each temperature in emulsion droplets containing three different concentrations of DNA.
We determine the volume fraction $x$ by assuming that the inner phase-separated droplet and the outer emulsion droplet are both spherical.
We then solve for the coexisting phase concentrations $\rho^{\text{L}}$ and $\rho^{\text{G}}$ as described below.
We also identify where each of these three total DNA concentrations intersects the binodal by locating the temperature at which the inner phase-separated droplets disappear; the uncertainty in these direct measurements of the binodal is less than $1^\circ$C.

\begin{figure}
  \begin{center}
    \includegraphics[width=\textwidth]{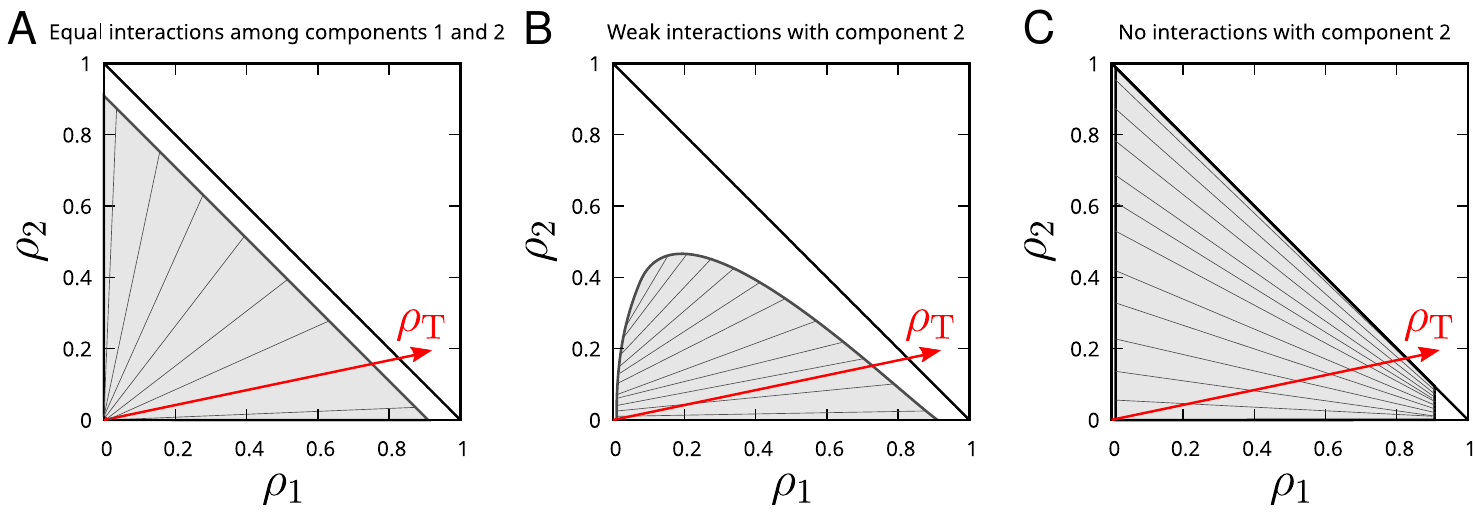}
  \end{center} \vskip-4ex
  \caption{Illustrative phase diagrams in a two-solute solution with (a) equal attractive interactions among the two components, (b) weaker interactions among solute 1 molecules and between solute 1 and 2 molecules, and (c) no net interactions among solute 1 molecule or between solute 1 and 2 molecules.  The two-phase coexistence region is shaded in light gray, and tie lines between coexisting phases on the binodal are indicated by thin dark gray lines.  We consider a mixture with a fixed composition, such that $\rho_2 / \rho_1 = \text{const}.$; varying the total solute concentration $\rho_{\text{T}} \equiv \rho_1 + \rho_2$ is equivalent to moving along the red vector shown in each panel.}
  \label{fig:phase_diagrams}
\end{figure}

According to \eqref{eq:lever}, the condensed-phase volume fraction $x$ should be a linear function of the total concentration, $\rho$, for $\rho^{\text{G}} \le \rho \le \rho^{\text{L}}$.
However, in our experiments, we generically find that $x(\rho)$ exhibits positive curvature.
Furthermore, the deviations from linearity tend to increase as the radii of the phase-separated droplets become small, indicating that the system is approaching the dilute-phase arm of the binodal.
These observations can be explained by considering how polydispersity affects the inferred binodal.
For simplicity, let us assume that there are two species present in the mixture: full-length strands (component 1) and partial-length strands (component 2), which comprise the minority species.
Partial-length strands are present in our experimental system as a result of premature termination during oligonucleotide synthesis.
To build intuition, we sketch schematic phase diagrams of two-solute-plus-solvent, constant-temperature phase diagrams in \figref{fig:phase_diagrams}.
In the limiting case on the left (\figref{fig:phase_diagrams}a), all the interactions (i.e., homotypic and heterotypic interactions) are the same, and thus the two components are physically equivalent.
In the limiting case on the right (\figref{fig:phase_diagrams}c), component 2 has no net interactions either with itself or with component 1; thus, this phase diagram is equivalent to a one-component phase diagram in which component 2 cannot be distinguished from the solvent.
Since we expect that partial-length strands interact weakly both with themselves and with the full-length strands, we anticipate that a phase diagram between these two limiting cases (\figref{fig:phase_diagrams}b) describes our experimental system.
Importantly, we note that except in the first limiting case, the tie lines are not parallel to a line of constant composition.

\begin{figure}
  \begin{center}
    \includegraphics[width=0.9\textwidth]{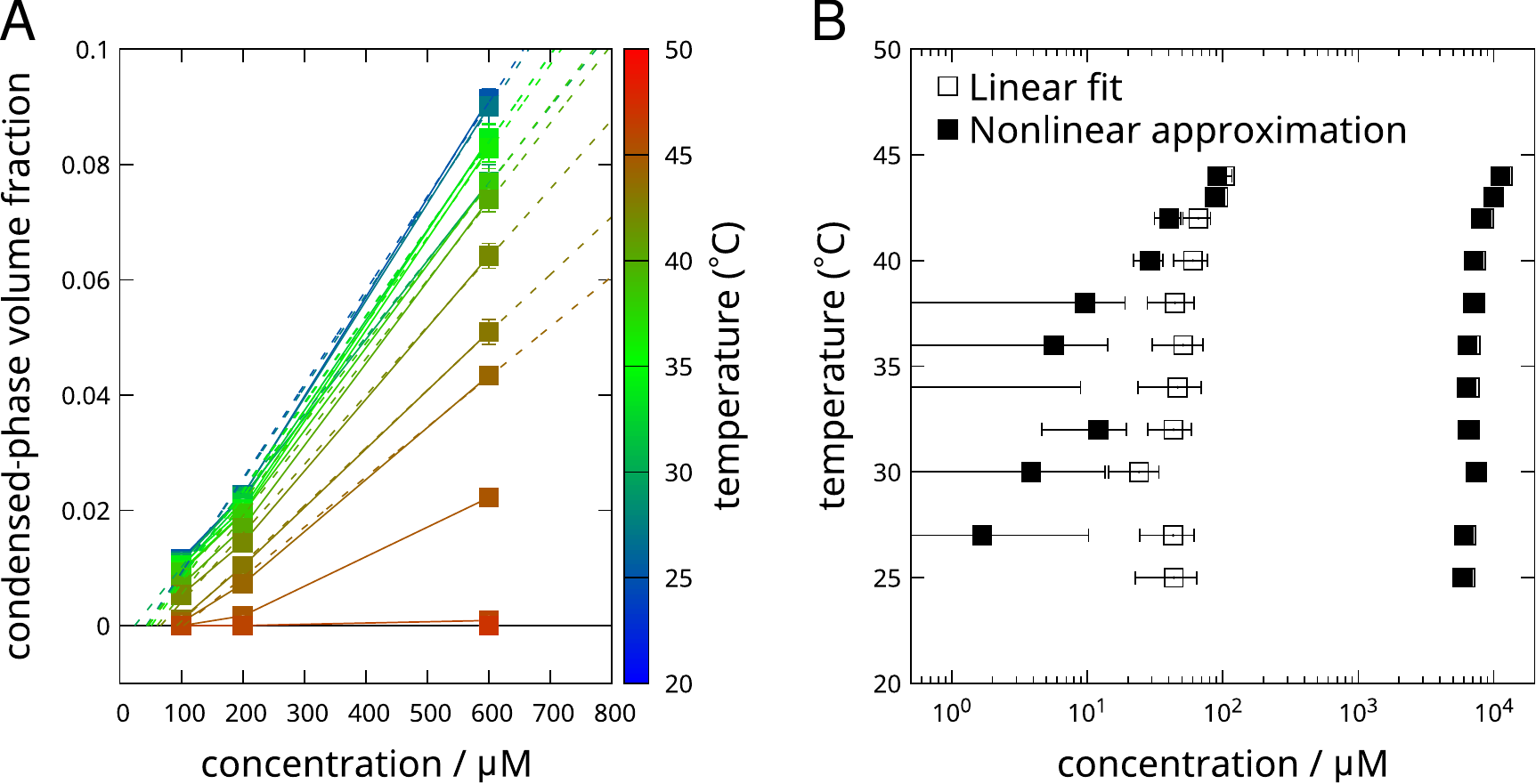}
  \end{center} \vskip-4ex
  \caption{(a)~Measurements of the condensed-phase volume fraction, $x$, at three total DNA concentrations, $\{\rho_{\text{T}}\}$, as a function of temperature (symbols and solid lines) for the low-temperature coexistence region of the with-hands system.  Comparison with a linear fit to the data at each temperature (dashed lines) reveals a slight yet consistent positive curvature of $x(\rho_{\text{T}})$.  This behavior is consistent with our theoretical analysis, \eqref{eq:lever-nonlinear}, which predicts positive curvature of $x(\rho_{\text{T}})$ when a fraction, $c$, of the strands/nanostars engage in weaker attractive interactions than the majority of the strands/nanostars in solution (such that $\Delta_{\text{G}}>0$ and $\Delta_{\text{L}}<0$; see text).  This scenario describes our experiments, since only $c \approx 25\%$ of the strands in solution are full length\footnote{The known per-nucleotide couple efficiency of the DNA oligonucleotides in our system is 0.994.  Since the with-hands sequences are 49 nucleotides long, the fraction of full-length strands is $\approx 0.994^{49} \approx 0.75$.} and shorter strands invariably interact more weakly, and are thus more probe to assembling defective nanostars, than full-length strands.  Consistent with this explanation, we observe the greatest deviations from the single-component lever rule in the vicinity of the UCST critical point (cf.~\figref{fig:phase_diagrams}b).
    (b)~Neglecting the positive curvature of $x(\rho_{\text{T}})$, and inferring the coexisting concentrations using \eqref{eq:lever} results in the ``linear fit'' phase diagram (open symbols).  By contrast, accounting for the nonlinearity of $x(\rho_{\text{T}})$ using local linear fits to the volume-fraction data, \eqsref{eq:lever-nonlinear-approx-dilute} and (\ref{eq:lever-nonlinear-approx-condensed}), results in the ``nonlinear approximation'' phase diagram (filled symbols).  The latter is more consistent with theoretical expectations, which predict that the concentration of the dilute phase decreases as the temperature is lowered (cf.~Fig.~2c in the main text).  Error bars indicate the uncertainties in the inferred concentrations of the coexisting phases due to statistical errors in the measurements of $x(\rho_{\text{T}})$ and linear extrapolation.}
  \label{fig:phase_diagrams_expt}
\end{figure}

To analyze how the presence of a weakly interacting component affects our experimental measurements, we consider a mixture of strands with partial-length mole fraction $c$, such that $\rho_2 / \rhot = c$ and $\rho_1/\rhot = 1 - c$, where $\rhot \equiv \rho_1 + \rho_2$ is the total strand concentration.
For small $c$, we can write the concentration of full-length strands on the binodal as
\begin{equation}
  \rho_1^{\text{G/L}} = \rho_1^{\text{G/L},0} + \Delta_{\text{G/L}} \times \rho_2,
\end{equation}
where $\rho_1^{\text{G/L},0}$ is the concentration of the gas/liquid phase in a pure solution containing component 1 only.
According to the weak-interaction phase diagram shown in \figref{fig:phase_diagrams}b, we assume that $\Delta_{\text{G}} > 0$ and $\Delta_{\text{L}} < 0$.
Writing the lever rule for $\rho_1$ in terms of $c$,
\begin{align*}
  \left(\rho_1^{\text{L},0} + \Delta_{\text{L}} \rhot c\right) x + \left(\rho_1^{\text{G},0} + \Delta_{\text{G}} \rhot c\right) (1 - x) &= (1 - c) \rhot \\
  \left[\rho_1^{\text{L},0} - \rho_1^{\text{G},0} + (\Delta_{\text{L}} - \Delta_{\text{G}}) \rhot c\right] x &= (1 - c - \Delta_{\text{G}} c) \rhot - \rho_1^{\text{G},0},
\end{align*}
we arrive at a nonlinear expression for the condensed-phase volume fraction $x(\rhot)$,
\begin{equation}
  \label{eq:lever-nonlinear}
  x(\rhot) = \frac{\left[1 - (1 + \Delta_{\text{G}}) c\right] \rhot - \rho_1^{\text{G},0}}{\rho_1^{\text{L},0} - \rho_1^{\text{G},0} + (\Delta_{\text{L}} - \Delta_{\text{G}}) c \rhot}.
\end{equation}
Thus, in a polymer solution with strand-length polydispersity (i.e., $c > 0$), the liquid-phase volume fraction should be a function of $\rhot$ with positive curvature, as observed in our experiments.
An analogous argument also applies to the nanostar UCST, in which case malformed nanostars comprise the weakly interacting minority species.
Example data for the ``with-hands'' nanostar system are shown in \figref{fig:phase_diagrams_expt}a.
Fundamentally, this nonlinearity arises from the fact that the system crosses tie lines as the total strand concentration is increased while the partial-length mole fraction, $c$, is held constant (\figref{fig:phase_diagrams}b).

In principle, we could infer the dilute and condensed-phase concentrations from measurements of the condensed-phase volume fractions at multiple total concentrations by fitting an equation of the form \eqref{eq:lever-nonlinear}, with four fitting parameters.
However, since we make measurements at only three total DNA concentrations, $\rhot^{(1)} < \rhot^{(2)} < \rhot^{(3)}$, we instead approximate the binodal by linear extrapolation of two secant lines.
Specifically, we estimate the dilute-phase concentration using
\begin{equation}
  \label{eq:lever-nonlinear-approx-dilute}
  \rhog \approx \frac{x^{(2)} \rhot^{(1)} - x^{(1)} \rhot^{(2)}}{x^{(2)} - x^{(1)}}
\end{equation}
and the condensed-phase concentration using
\begin{equation}
  \label{eq:lever-nonlinear-approx-condensed}
  \rhol \approx \frac{(1 - x^{(2)}) \rhot^{(3)} - (1 - x^{(3)}) \rhot^{(2)}}{x^{(3)} - x^{(2)}}.
\end{equation}
A comparison of phase diagrams inferred using \eqref{eq:lever} versus \eqsref{eq:lever-nonlinear-approx-dilute} and (\ref{eq:lever-nonlinear-approx-condensed}) is shown in \figref{fig:phase_diagrams_expt}b.
The uncertainty in these estimates is calculated by computing the standard error of the volume-fraction measurements, $x^{(1)}$, $x^{(2)}$, and $x^{(3)}$, and performing a linear error propagation.
Due to the positive curvature of \eqref{eq:lever-nonlinear}, these estimates should be interpreted as upper bounds on the true values of $\rhog$ and $\rhol$.

\section{Theoretical model}

In this section, we summarize the theoretical model and present analytical expressions for the predicted binodals in relevant limits.
For a more detailed derivation of the model and a discussion of its phase behavior in different parameter regimes, we refer the reader to Ref.~\cite{li2023interplay}.

\begin{figure}[h]
    \centering
    \includegraphics[width=0.6\columnwidth]{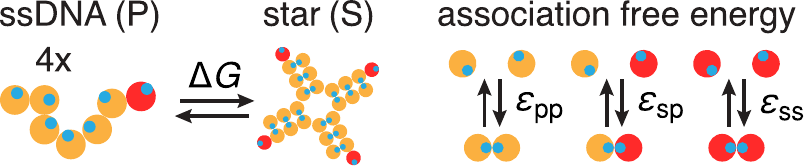}
    \caption{Schematic of the mean-field model.  Disassembled ssDNA strands (P), consisting of $N_{\text{p}}$ blobs (yellow circles), self-assemble into a nanostar (S) in a two-state reaction.  Hand-shaking blobs (red circles) are included if present in a particular system.  Blobs interact via the association of one-to-one binding sites (small blue circles) according to the association free energies $\epsilon_{\text{pp}}$, $\epsilon_{\text{ps}}$, and $\epsilon_{\text{ss}}$.}
    \label{fig:model-schematic}
\end{figure}

\subsection{Free energy density}

In our model (\figref{fig:model-schematic}), we consider an assembled nanostar to be a rigid particle and a DNA strand to be a freely jointed polymer chain consisting of $N_{\text{p}}$ blobs. The volume of a blob, $v_0$, is the Kuhn volume of single-stranded DNA. The two components in the idealized mixture are the nanostar (S) and single-stranded DNA (P), along with an implicit solvent. S and P interconvert through the chemical reaction $n \text{P} \rightleftharpoons \text{S}$, while mass conservation holds for the DNA strands,
\begin{equation}
\label{eq:mass}
    \rho_{\text{p}} + n\rho_{\text{s}} = \rho_{\text{total}}.
\end{equation}

We break down the free energy density of the mixture into three parts: an internal energy contribution, a purely entropic contribution due to mixing, and the association free energy due to hybridization:
\begin{equation}
\label{eq:fedensity}
    f = f^{\text{int}}+f^{\text{EFH}}+f^{\text{A}}.
\end{equation}
The internal energy contribution is
\begin{equation}
    \beta f^{\text{int}} = \rho_{\text{s}} \beta u_{\text{s}},
\end{equation}
where $\rho_{\text{s}}$ is the number density of S.
By setting $u_{\text{p}}=0$, $u_{\text{s}}$ represents the internal energy difference between S and P.
Next, we account for the ideal and excluded-volume contributions to the free energy density of the nanostar/polymer mixture using an ``extended Flory--Huggins'' expression~\cite{li2023interplay}.
\begin{equation}
  \beta f^{\text{EFH}} v_0 = \frac{\phi_{\text{s}}}{N_{\text{s}}} \ln \phi_{\text{s}} + \frac{1 -
  \phi_{\text{s}}}{N_{\text{s}}} \ln (1 - \phi_{\text{s}}) + \frac{\phi_{\text{p}}}{N_{\text{p}}} \ln \left(
  \frac{\phi_{\text{p}}}{1 - \phi_{\text{s}}} \right) + (1 - \phi_{\text{s}} - \phi_{\text{p}}) \ln \left( \frac{1 - \phi_{\text{s}} - \phi_{\text{p}}}{1 - \phi_{\text{s}}} \right),
\end{equation}
where $\phi_{\text{s}} \equiv N_{\text{s}}v_0\rho_{\text{s}}$ is the volume fraction occupied by nanostars and $\phi_{\text{p}} \equiv N_{\text{p}}v_0\rho_{\text{p}}$ is the volume fraction occupied by polymer.
$N_{\text{p}}$ and $N_{\text{s}}$ are the number of blobs per polymer and the number of blob volumes corresponding to the excluded volume of an assembled nanostar, respectively.

We employ statistical associating fluid theory (SAFT)~\cite{chapman1989saft} to describe the free-energy contribution arising from strand hybridization.
Using an ideal lattice gas as the reference state \cite{jacobs2014phase}, the theory expresses the associative Helmholtz free energy density $f^{\text{A}}$ in the form
\begin{equation}
  \beta f^{\text{A}} = \sum_i \rho_i \sum_A \left( \ln X_{iA} -
  \frac{X_{iA}}{2} + \frac{1}{2} \right),
\end{equation}
where $X_{iA}$ is the unbound probability of a type-A binding site on species $i$ at equilibrium and $\rho_i$ is the number density of species $i$.
At chemical equilibrium, the unbound probabilities can be solved self-consistently from the system of equations
\begin{equation}
  X_{iA} = \frac{1}{1 + v_0 \sum_j \rho_j \sum_B X_{jB} \alpha_{iA,jB} e^{-\beta\epsilon_{iA,jB}}},
  \label{SAFTx}
\end{equation}
where $\epsilon_{iA,jB}$ is the dimensionless binding strength between binding site $A$ on species $i$ and binding site $B$ on species $j$.
The factors $\alpha_{iA,jB}$ reflect geometric constraints on these interactions, while $\epsilon_{iA,jB}$ is related to the equilibrium constant $K_{iA,jB}$~\cite{michelsen2001physical} for forming a dimer between a pair of Kuhn length-sized DNA subsequences~\cite{jacobs2015rational},
\begin{equation}
  -\beta\epsilon_{iA,jB} = \ln \left[(1 + \delta_{iA,jB}) 
  K_{iA,jB} v^{- 1}_0 \right].
  \label{Ktoepsilon}
\end{equation}

In our model, one of the binding strengths is associated with complementary hand-shaking interactions, $\epsilon_{\text{ss}}$, between assembled nanostars.
Each nanostar has $n=4$ binding sites of this type (indicated by the subscript s).
For simplicity, we assume that all polymer--polymer hybridization interactions can be captured by a single average binding strength, $\epsilon_{\text{pp}}$, and that the cross interactions between nanostar hand sequences and polymer blobs can be represented by $\epsilon_{\text{sp}}$ (\figref{fig:parameterization}a).
Each polymer thus has $N_{\text{p}}$ equivalent binding sites.
With these approximations, the associative free energy density takes the form
\begin{equation}
  \label{eq:fedensity-assoc}
  \beta f^{\text{A}} = \rho_{\text{p}} N_{\text{p}} \left( \ln X_{\text{p}} - \frac{X_{\text{p}}}{2} + \frac{1}{2}
  \right) + \rho_{\text{s}} n \left( \ln X_{\text{s}} - \frac{X_{\text{s}}}{2} + \frac{1}{2} \right).
\end{equation} 
Hand-shaking interactions between nanostars are constrained by the solid angle $\Omega$ spanned by each ``hand'' blob (\figref{fig:parameterization}b), such that $\alpha_{\text{ss}} = \Omega / 4\pi$.
However, there are no such constraints for the nanostar--polymer interactions and polymer--polymer interactions, so $\alpha_{\text{sp}} = \alpha_{\text{pp}} = 1$.
The self-consistent chemical equilibrium equations for the s and p binding sites in our model are thus
\begin{align}
  X_{\text{s}} &= \frac{1}{1 + v_0 \rho_{\text{s}} n X_{\text{s}} \alpha_{\text{ss}} e^{-\beta\epsilon_{\text{ss}}} + v_0 \rho_{\text{p}} N_{\text{p}} X_{\text{p}} e^{-\beta\epsilon_{\text{sp}}}},
  \label{eq:Xs} \\
  X_{\text{p}} &= \frac{1}{1 + v_0 \rho_{\text{s}} n X_{\text{s}} e^{-\beta\epsilon_{\text{sp}}} + v_0 \rho_{\text{p}} N_{\text{p}} X_{\text{p}} e^{-\beta\varepsilon_{\text{pp}}}}.
  \label{eq:Xp}
\end{align}

\subsection{Chemical equilibrium}

Chemical equilibrium between the polymer (P) and nanostar (S) states requires that
\begin{equation}
  \label{eq:chemeq}
  n \mu_{\text{p}} = \mu_{\text{s}},
\end{equation}
where $n=4$ strands comprise each nanostar in our experimental system.
Following from \eqref{eq:fedensity}, the chemical potential of each species can be written
\begin{align}
  \beta\mu_{\text{p}} &= \ln\phi_{\text{p}} + (1 - N_{\text{p}}) + (N_{\text{p}} - 1) \ln (1 - \phi_{\text{s}}) - N_{\text{p}} \ln (1 - \phi_{\text{s}} - \phi_{\text{p}}) + N_{\text{p}} \ln X_{\text{p}}, \\
  \beta\mu_{\text{s}} &= \ln\phi_{\text{s}} + \beta u_{\text{s}} - \ln (1 - \phi_{\text{s}}) + N_{\text{s}} \ln \left( \frac{1 - \phi_{\text{s}}}{1 - \phi_{\text{s}} - \phi_{\text{p}}} \right) - N_{\text{s}} \left( 1 - \frac{1}{N_{\text{p}}} \right)\frac{\phi_{\text{p}}}{1 - \phi_{\text{s}}} + n \ln X_{\text{s}}.
\end{align}
By applying the chemical equilibrium condition, \eqref{eq:chemeq}, at dilute concentrations, we can relate the internal energy difference, $u_{\text{s}}$, to the free-energy of nanostar formation, $\Delta G$,
\begin{equation}
  \beta u_{\text{s}} = \beta \Delta G + n(1-N_{\text{p}}).
  \label{eq:deltaG}
\end{equation}

\subsection{Phase coexistence calculations}

At thermodynamic equilibrium, the system reaches both chemical equilibrium and phase equilibrium \cite{Zwicker_2022,bauermann2022chemical}. Phase equilibrium requires equal pressures and chemical potentials across all coexisting phases. For a free energy function with a single concentration variable, the coexisting phases can be determined by the common tangent construction \cite{rubinstein2003polymer}, where the DNA concentrations in the two coexisting phases are given by the tangential points on the free energy landscape. The common tangent construction ensures equal pressures and chemical potentials in both phases.

At a specific total DNA strand concentration $\rho_{\text{total}}$, we first solve for $\rho_{\text{s}}$ and $\rho_{\text{p}}$ at chemical equilibrium, \eqref{eq:chemeq}, while enforcing mass conservation, \eqref{eq:mass}. We then calculate the free energy density, \eqref{eq:fedensity}, at  $\rho_{\text{total}}$ accordingly. If there are multiple solutions to \eqref{eq:chemeq}, we take the one with the lowest free energy density. In this way, we obtain the free energy landscape $f(\rho_{\text{total}})$. Lastly, we find the common tangents by calculating the convex hull of a discretized set of points on the free energy landscape \cite{mao2019phase}.

\subsection{Analytical results for phase boundaries}

To gain insight into the determinants of the high-temperature binodal, we derive analytical formulae for the phase boundaries in two limits: $\beta \Delta G \rightarrow \infty$ and $\beta \Delta G \rightarrow - \infty$.
Specifically, by considering the nanostar system without hand sequences, we aim to obtain approximate expressions relating $\beta \Delta G$ and $\beta \epsilon$ at a specified dilute-phase concentration $\rho_{\text{v}}$. 

At phase coexistence, both phases have identical pressures and chemical potentials
\begin{align}
  P^{\text{v}} &= P^{\text{l}}, \\
  \mu^{\text{v}} &= \mu^{\text{l}}.
  \label{equal_chem}
\end{align}
We use the superscript v for the dilute phase, and l for the condensed phase.
The pressure and chemical potential are calculated from the free energy \eqref{eq:fedensity} via
\begin{align}
    P &= f - \mu \rho_{\text{total}}, \\
    \mu &= \frac{\partial f}{\partial \rho_{\text{total}}},
\end{align}
where $\rho_\text{total}$ is the concentration of DNA strands, as given by \eqref{eq:mass}.

We assume that the coexistence pressure is $P \approx 0$.
This assumption is valid when the dilute phase can be treated as an ideal gas, as the DNA concentration is very low.
For both limits of $\beta \Delta G$, the condensed phase is dominated by the polymer (P) state.
We therefore assume that the dense phase is 100\% polymer and that $\phi^{\text{l}} e^{- \beta \epsilon_{\text{pp}}} \gg 1$.
In this strong-binding limit, the unbound probability from \eqref{SAFTx} is
\begin{equation}
  X_{\text{p}}^{\text{l}} = (\phi_{\text{p}}^{\text{l}} e^{- \beta \epsilon_{\text{pp}}})^{- \frac{1}{2}}.
\end{equation}
Thus, we have
\begin{equation}
  - \beta P^{\text{l}} v_0 = \ln (1 - \phi^{\text{l}}) - (1 - N_{\text{p}})
  \frac{\phi^{\text{l}}}{N_{\text{p}}} + \frac{1}{2} \phi^{\text{l}} = 0
  \label{liquid_density}
\end{equation}
at phase coexistence.
 Solving for $\phi^{\text{l}} \equiv \rho^{\text{l}} N_{\text{p}}v_0$ using \eqref{liquid_density}, we find that $\phi^{\text{l}} \approx 0.454$ for $N_{\text{p}} = 6$.

The chemical potential in the condensed phase is
\begin{equation}
  \beta \mu^{\text{l}} = \left( 1 - \frac{N_{\text{p}}}{2} \right) \ln \phi^{\text{l}}
  - N_{\text{p}} \ln (1 - \phi^{\text{l}}) + \frac{N_{\text{p}}}{2} \beta \epsilon + 1 - N_{\text{p}},
\end{equation}
while in the dilute phase, we assume that only the ideal and internal-energy terms contribute to the chemical potential since $\phi_{\text{v}}\ll 1$. Next, we discuss the equal-chemical-potential condition in both limits of $\beta \Delta G$.

\begin{figure}
    \centering
    \includegraphics[width=0.6\columnwidth]{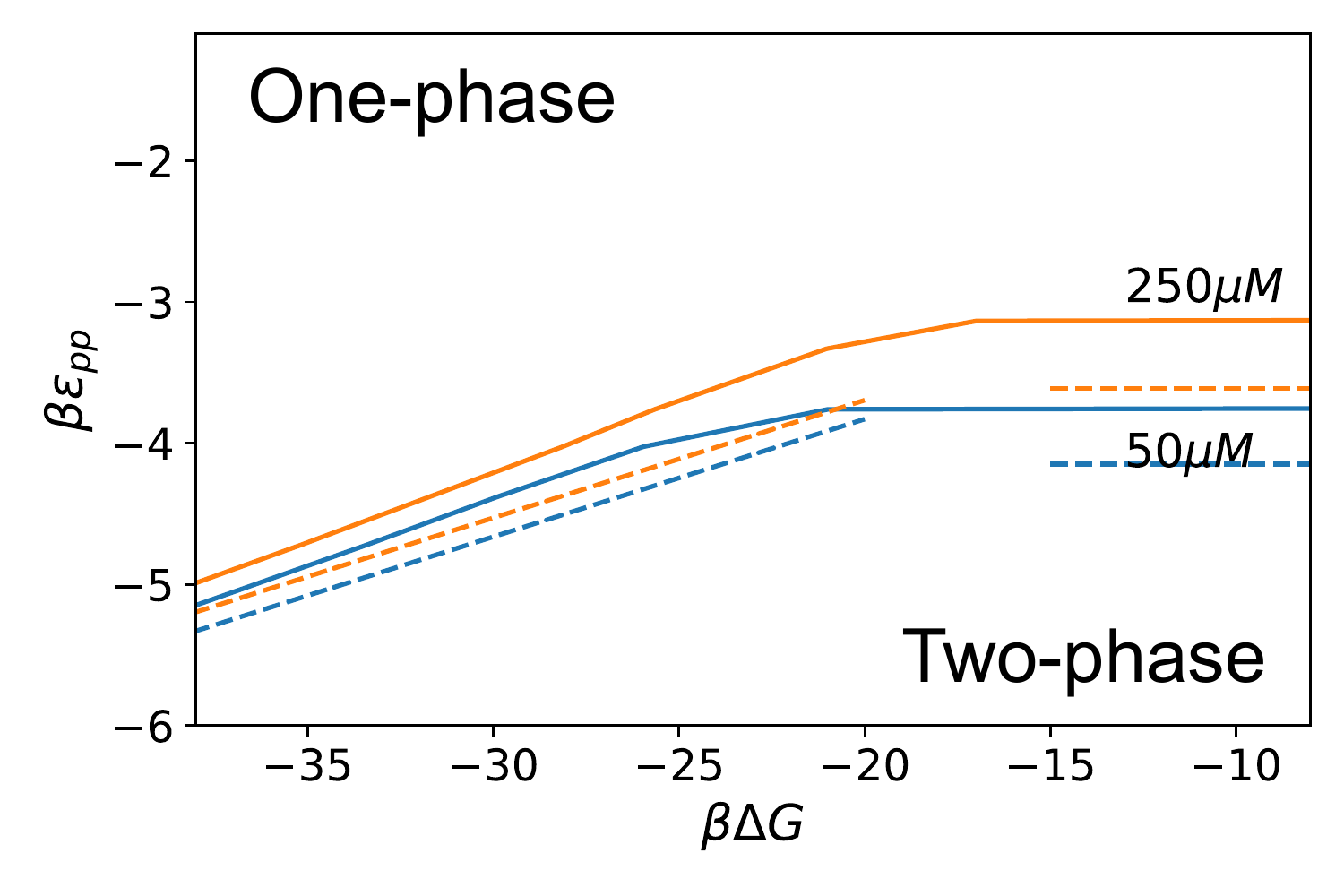} \vskip-1.5ex
    \caption{Predicted phase boundaries at two values of the total DNA strand concentration. Dashed lines show analytical results in both the $\beta \Delta G \rightarrow -\infty$ and $\beta \Delta G \rightarrow \infty$ limits.}
    \label{fig:phase_boundary}
\end{figure}

\subsubsection{Stable nanostar limit $\beta \Delta G \rightarrow -\infty$}

In this limit, the nanostars are extremely stable, and so we assume that the dilute phase is $100\%$ nanostar.
The chemical potential in the dilute phase is
\begin{equation}
  \beta \mu^{\text{v}} = \frac{1}{n} [\ln \phi^{\text{v}} + \beta \Delta G + n (1 - N_{\text{p}})],
\end{equation}
where $\phi^{\text{v}} = \rho^{\text{v}} N_{\text{s}} v_0 / n$.
By enforcing equal chemical potentials for both phases, \eqref{equal_chem}, we arrive at Eq.~(2) in the main text:
\begin{equation}
    \beta\Delta G - \ln\left(\frac{n}{\rho^{\text{v}} v_{\text{s}}}\right) = \frac{n N_{\text{p}} \beta\epsilon_{\text{pp}}}{2} - \ln \left[ \frac{(1 - \phi^{\text{l}})^{nN_{\text{p}}}}{(\phi^{\text{l}})^{n(1-N_{\text{p}}/2)}} \right].
\end{equation}
This equation predicts a linear relation between $\beta \Delta G$ and $\beta \epsilon_\text{pp}$ for a given dilute phase concentration $\rho^\text{v}$ (\figref{fig:phase_boundary}).

\subsubsection{Unstable nanostar limit $\beta \Delta G \rightarrow \infty$}

In this limit, the nanostar is extremely unstable, and so we assume that both the dilute phase and the condensed phase are $100\%$ polymer.
The chemical potential in the dilute phase is
\begin{equation}
  \beta \mu^{\text{v}} = \ln \phi^{\text{v}} + 1 - N_{\text{p}}.
\end{equation}
In this expression, $\phi^{\text{v}} =\rho^{\text{v}}N_{\text{p}} v_0$.
We again enforce equal chemical potentials for both phases \eqref{equal_chem}:
\begin{equation}
  \frac{N_{\text{p}}}{2} \beta \epsilon_\text{pp} = \ln (\rho^\text{v} N_\text{p} v_0) + N_\text{p} \ln (1 -
  \phi^\text{l}) - \left( 1 - \frac{N_\text{p}}{2} \right) \ln \phi^\text{l}.
\end{equation}
The dilute phase concentration $\rho^{\text{v}}$ is entirely determined by $\beta \epsilon_\text{pp}$ and $N_\text{p}$ in this limit (\figref{fig:phase_boundary}).

\section{Model parameterization}

In order to make predictions for a specific experimental system, it is necessary to specify the parameters in the free energy density described above.
Here we describe how the parameters were obtained for computing the theoretical phase diagrams presented in the main text.
We then discuss the sensitivity of our predicted phase diagrams to these parameter choices.

\subsection{Parameterization approach}

We first estimate the values of the excluded volume parameters, including the Kuhn segment volume $v_0$ and the excluded volume of both species, $N_{\text{s}}$ and $N_{\text{p}}$, in units of $v_0$, on the basis of previous experimental results \cite{CHI20131072}.
A Kuhn segment of a ssDNA is about 7-nt long, occupying a volume of $v_0 = 4 {\text{nm}}^3$ (or $v^{- 1}_0 = 0.415 \text{M}$).
Each DNA strand without hands (see Fig.~1a in the main text) is 43-nt long and thus consists of six blobs ($N_{\text{p}}=6$).
DNA strands with hands are 49-nt long and thus consist of seven blobs ($N_{\text{p}}=7$), where the last blob constitutes one hand of a nanostar.
We estimate the excluded volume of a nanostar to be $v_{\text{s}} = N_{\text{s}} v_0$, where $N_{\text{s}}=200$, on the basis of the experimentally determined nanostar concentration in the condensed phase at low temperatures (see Fig.~2c in the main text).

\begin{figure}[t]
  \includegraphics[width=\columnwidth]{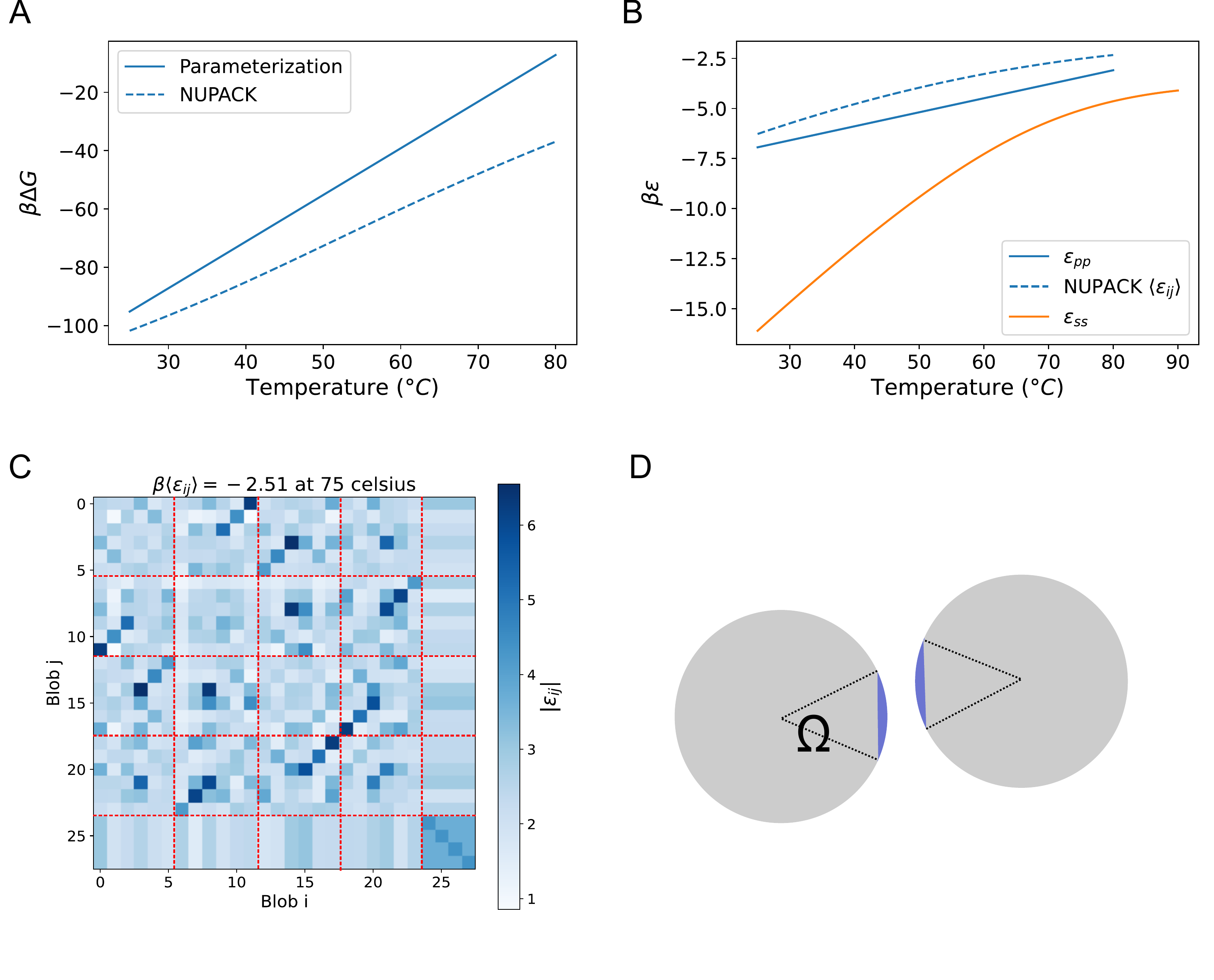}
  \caption{(a) The free energy of nanostar formation, $\Delta G$, as a function of temperature.
    (b)~Our parameterization of the average blob hybridization free energy, $\epsilon_\text{pp}$, is close to the predictions of NUPACK~\cite{NUPACK2020}. We directly use the value for the binding free energy between two hand subsequences, $\epsilon_\text{ss}(T)$, that is predicted by NUPACK.
    (c)~A representative dimensionless binding strength matrix, $\beta\varepsilon_{ij}$, between subsequences $i$ and $j$ calculated from NUPACK at $75^\circ\text{C}$. Starting from index 0, each set of six consecutive indices belongs to the same DNA strand; the last four rows/columns represent the interactions between the hand subsequence and all other subsequences.
    (d)~Hand-shaking interactions are constrained by the solid angle $\Omega$ (see text).}
  \label{fig:parameterization}
\end{figure}

The main challenge for parameterization is to determine the blob hybridization free energies $\epsilon_{\text{pp}}(T)$, $\epsilon_{\text{sp}}(T)$ and $\epsilon_{\text{ss}}(T)$ and the free energy of nanostar formation $\Delta G(T)$ as functions of temperature (\figref{fig:parameterization}a,b).
For the nanostar system without hand subsequences, we have $\epsilon_{\text{sp}}(T) = \epsilon_{\text{pp}}(T) = 0$.
As noted in the main text, we assume that $\Delta G$ is linearly dependent on the $\epsilon_{\text{pp}}$.
We further assume that both $\Delta G$ and $\epsilon_{\text{pp}}$ are linear functions of the absolute temperature, $T$.
This assumption implies that each of these free energies can be written, approximately, in the form $H^\circ - T S^\circ$, where $H^\circ$ and $S^\circ$ represent temperature-independent enthalpic and entropic contributions, respectively~\cite{santalucia2004thermodynamics}.
Within the constraints implied by these assumptions, we fit a temperature-dependent line on the ``master phase diagram'' (Fig.~3b in the main text) that most closely reproduces the experimental phase diagram for the no-hands system (Fig.~2b in the main text).
We use the same parameterization for $\epsilon_{\text{pp}}(T)$ for all systems studied because we expect the average blob hybridization free energy to be relatively insensitive to point mutations or the addition of hand subsequences.

For the system with hand subsequences, we estimate $\epsilon_{\text{ss}}(T)$ according to the nanostar upper critical solution temperature (UCST) measured in the experiments.
We first obtain the binding free energy between two complimentary hand subsequences, $\epsilon_\text{ss}(T)$, directly from NUPACK~\cite{NUPACK2020} (\figref{fig:parameterization}b).
We then set the geometric factor $\alpha_{\text{ss}}=\nicefrac{1}{16}$ in \eqref{eq:Xs} to match the nanostar UCST measured in experiments.
We note that this value is comparable to the what we should expect on the basis of geometrical considerations, $\alpha_{\text{ss}} \simeq N_{\text{s}}^{-2/3} \approx \nicefrac{1}{34}$ (\figref{fig:parameterization}d).
Since $\epsilon_{\text{sp}}(T)$ plays a minor role in determining the phase behavior, we set $\epsilon_{\text{sp}}(T)=\epsilon_{\text{pp}}(T)$ in the with-hands system.
We further assume that the addition of hand subsequences has little effect on the nanostar stability and thus use the same parameterization for $\Delta G(T)$ in both the with-hands and no-hands systems.

Lastly, we modify the free energy of nanostar formation, $\Delta G(T)$, due to point mutations, which tend to destabilize the nanostars.
We assume that the point mutations are independent and have equivalent effects on the free energy of nanostar formation, such that
\begin{equation}
  \Delta G(T;m) = \Delta G(T;0) + nm\Delta\Delta G,
\end{equation}
where $m$ is the number of mutations per strand.
We find that a per-bulge free-energy penalty of ${\Delta\Delta G=0.95\,k_\text{B} T}$ provides the best fit to the transition temperatures of the mutants at a total DNA concentration of $200\,\mu \text{M}$.

\subsection{Validation of parameters}

\subsubsection{Blob hybridization free energy}

In this section, we show that our parameterization for $\epsilon_{\text{pp}}(T)$ is compatible with the predictions of NUPACK~\cite{NUPACK2020}.
NUPACK predicts equilibrium constants for DNA hybridization by calculating partition functions, which account for partial and complete hybridization of pairs of DNA strands.
The ensemble partition function $Q(\psi)$ of nucleic acid sequence(s) $\psi$ sums over the Boltzmann factors of secondary structures $\{x\}$,
\begin{equation}
  Q (\psi) = \sum_x e^{- \beta G (\psi ; x)}.
\end{equation}
For a simple reaction $A + B \rightleftharpoons AB$, the equilibrium constant $K_{AB}$ can be directly computed from these partition functions,
\begin{equation}
  K_{AB} = \frac{Q_{AB}}{Q_A Q_B \rho_{H_2 O}},
\end{equation}
where the constant $\rho_{\text{H}_2\text{O}} = 55.14\,\text{M}$ is the reference concentration for calculating partition functions in aqueous solution.
From these equilibrium constants, we compute dimensionless binding strengths using \eqref{Ktoepsilon}.
Specifically, each dimensionless binding strength $\epsilon_{iA,jB}$ in \eqref{SAFTx} represents the binding free energy between a pair of blobs, including the symmetry factor $(1 + \delta_{iA,jB})$ in \eqref{Ktoepsilon}.

In the with-hands system, there are 28 total 7-nt subsequences, of which 25 are unique.
The dimensionless binding free energies between subsequences form a $28\times28$ matrix (\figref{fig:parameterization}c).
In systems without hand sequences, the binding strength matrix has dimensions $24\times24$.
In accordance with our mean-field model, which does not distinguish among the polymer blobs for simplicity, we do not explicitly consider heterogeneity in the polymer-polymer binding strengths.
Instead, we consider the mean binding strength $\langle \epsilon_{ij}\rangle$, where $\langle\cdot\rangle$ indicates an average over the entire blob interaction matrix (\figref{fig:parameterization}c).
Importantly, we find that our parameterization of $\epsilon_{\text{pp}}(T)$ is very similar to the NUPACK prediction for $\langle \epsilon_{ij}\rangle$, except that our parameterization is approximately $1\, k_{\text{B}} T$ lower than $\langle \epsilon_{ij}\rangle$ (\figref{fig:parameterization}b).
This favorable comparison provides further support for our hypothesis that hybridization between blob-length subsequences contributes significantly to the associative interactions among disassembled DNA at high temperatures.

We note that various sources contribute to the uncertainty in $\varepsilon_{iA,jB}$ estimated from NUPACK.
First, NUPACK is parameterized at salt concentrations and temperatures that differ substantially from the current experimental conditions.
Second, in our system, the blob subsequences are parts of longer chains, while the NUPACK calculations are performed assuming pairs of isolated 7-nt subsequences.
Third, NUPACK does not explicitly consider the three-dimensional structure of the strands.
Finally, as noted above, we assume for simplicity that the heterogeneity in the binding strengths seen in \figref{fig:parameterization}c can be ignored.
These uncertainties may explain the minor differences between our parameterization of $\epsilon_{\text{pp}}(T)$ and the NUPACK prediction for $\langle \epsilon_{ij}\rangle$.

\subsubsection{Free energy of nanostar formation}

The dimensionless quantity $\beta \Delta G$ is the internal free energy change of nanostar formation, resulting from the reaction $n P \rightleftharpoons S$ in the dilute limit.
This quantity is related to the fraction of strands that are hybridized (or `bound'), $\xi$, in dilute-concentration melting experiments. In an ideal solution with strand concentration $\rho$, $\xi$ is related to the free energy in the model $\beta \Delta G$ \eqref{eq:deltaG} through
\begin{equation}
    \beta \Delta G = -\ln\left[ \frac{\xi}{4(\rho/\rho^\circ)^{n-1}(1-\xi)^n}\right] - \ln\left(v_0 \rho^\circ N_{\text{s}}\right) + n \ln \left(v_0 \rho^\circ N_{\text{p}}\right),
\end{equation}
where $\rho^\circ = 1$M is the commonly used reference concentration. In the melting experiments, we have total strand concentration $\rho = 1\mu$M, and the number of strands per nanostar $n=4$.

\begin{figure}[t]
  \includegraphics[width=\columnwidth]{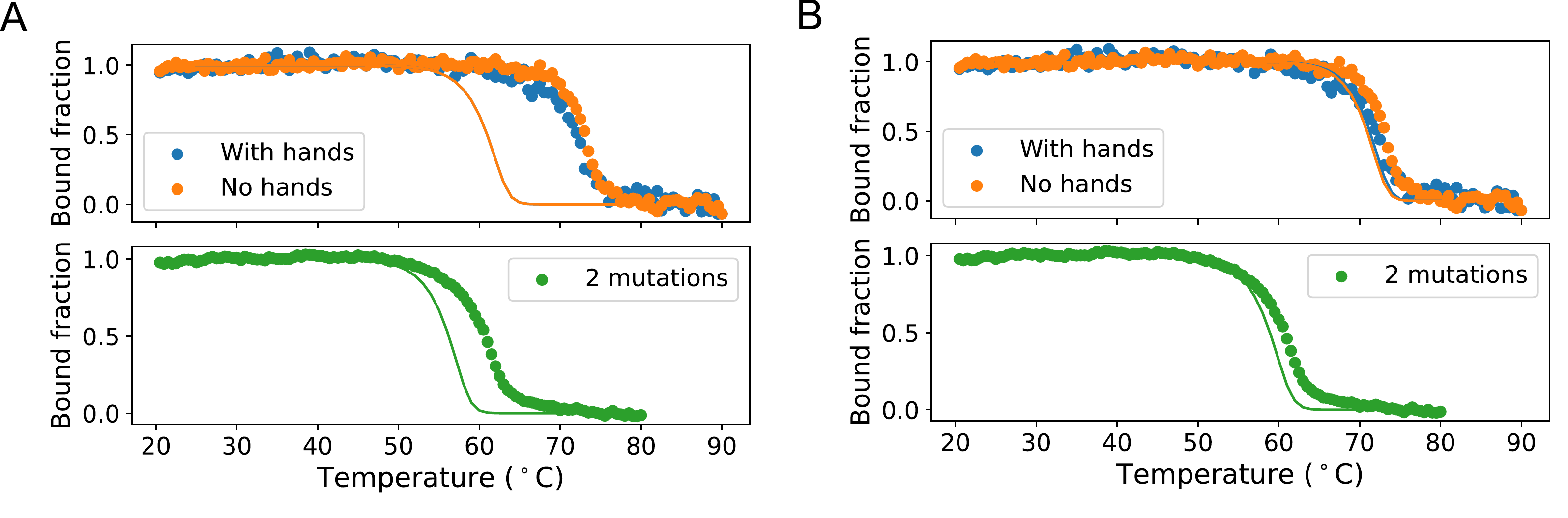}
  \caption{(a)~Validation of the $\beta\Delta G(T)$ parameters by comparison of predicted bound fractions in the two-state approximation (solid lines) to experimental melting curves (dots) at 1 $\mu$M DNA. \textit{Top:} The same $\beta \Delta G(T)$ parameterization is used for both the with-hands and no-hands systems.  \textit{Bottom:} Melting curves from experiments and the  $\beta \Delta G(T;m=2)$ parameterization used in the main text for the 2-bulge-per-arm construct. (b)~The alternative parameterization described in SI~\secref{sec:alt-param}.}
  \label{fig:melting-validation}
\end{figure}

As described in the main text, we choose $\beta \Delta G(T)$ to fit the phase diagram of nanostars without hands (\figref{fig:parameterization}a).
This parameterization, which is close to the prediction obtained from NUPACK (\figref{fig:parameterization}a), results in a slight discrepancy with the experimentally determined melting curves at 1$\mu$M DNA (\figref{fig:melting-validation}a).
This discrepancy most likely arises from the simplified two-state model of nanostar assembly, which does not account for intermediate states that contribute to the bound fraction at higher temperatures.

\subsection{Alternative parameterization}
\label{sec:alt-param}
\begin{figure}
  \includegraphics[width=\columnwidth]{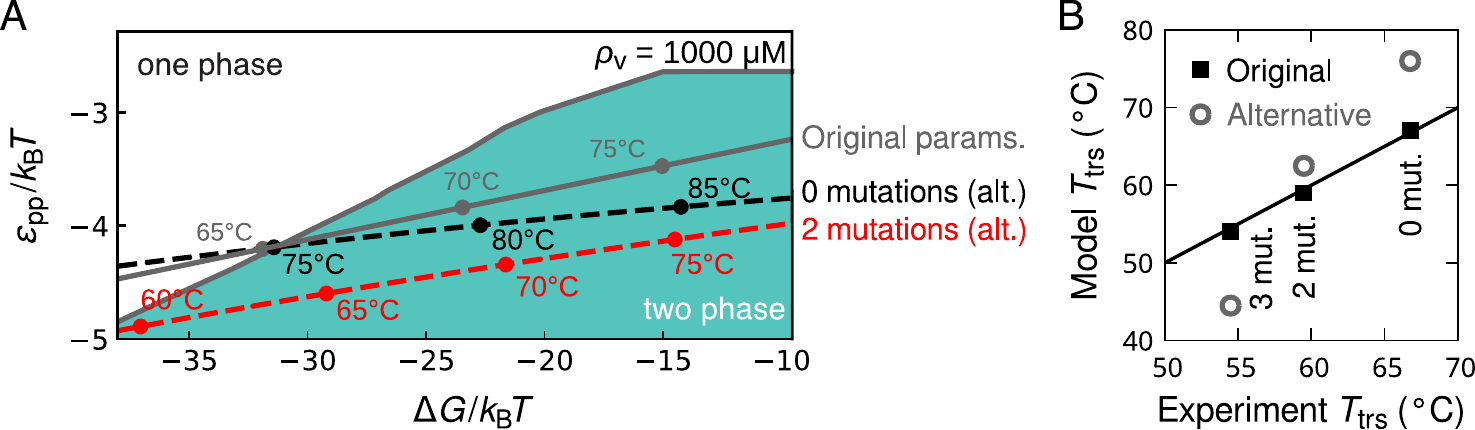}
  \caption{An alternative parameterization of the temperature-dependent model parameters $\beta\Delta G$ and $\beta\epsilon_{\text{\text{pp}}}$ yields qualitatively similar predictions.
    (a)~Dashed curves show the alternative parameterization that reproduces the experimental melting curves shown in \figref{fig:melting-validation}b.
    (b)~Both parameterizations predict decreasing transition temperatures at a total DNA concentration of 200 $\mu $M as mutations are introduced, although the transition temperature is more sensitive to the number of mutations in the alternative parameterization (circles) than in the model parameterization presented in the main text (squares; cf.~Fig.~4 in the main text).}
  \label{fig:alt-param}
\end{figure}

In this section, we consider an alternative parameterization that attempts to reproduce the melting curves quantitatively using our two-state approximation for nanostar assembly  (\figref{fig:alt-param}a).
To this end, we first determine $\beta \Delta G(T)$ for the with-hands and no-hand systems by fitting to the experimental melting curve (\figref{fig:melting-validation}b). Similarly, we determine $\beta \Delta G(m=2,T)$ by fitting the melting curve of 2-mutation-per-strand nanostars. We then extract the free energy penalty of a ``bulge" using $\Delta\Delta G(T)= [\Delta G(T;m=2)-\Delta G(T;m=0)] / 2$.
Finally, we set $\epsilon_{\text{pp}}(T)$ equal to the NUPACK prediction, $\langle \epsilon_{ij}\rangle$, plus an empirical additive constant of $(-\ln 5) k_{\text{B}}T$.
Using this parameterization, our theoretical model still makes qualitatively correct predictions for the phase diagrams (\figref{fig:alt-param}b).
This analysis demonstrates that the conclusions described in the main text are insensitive to the specific parameterization as long as they are in a physically reasonable range.

\section{Effect of changing the DNA strand length}

\begin{figure}
  \includegraphics[width=\columnwidth]{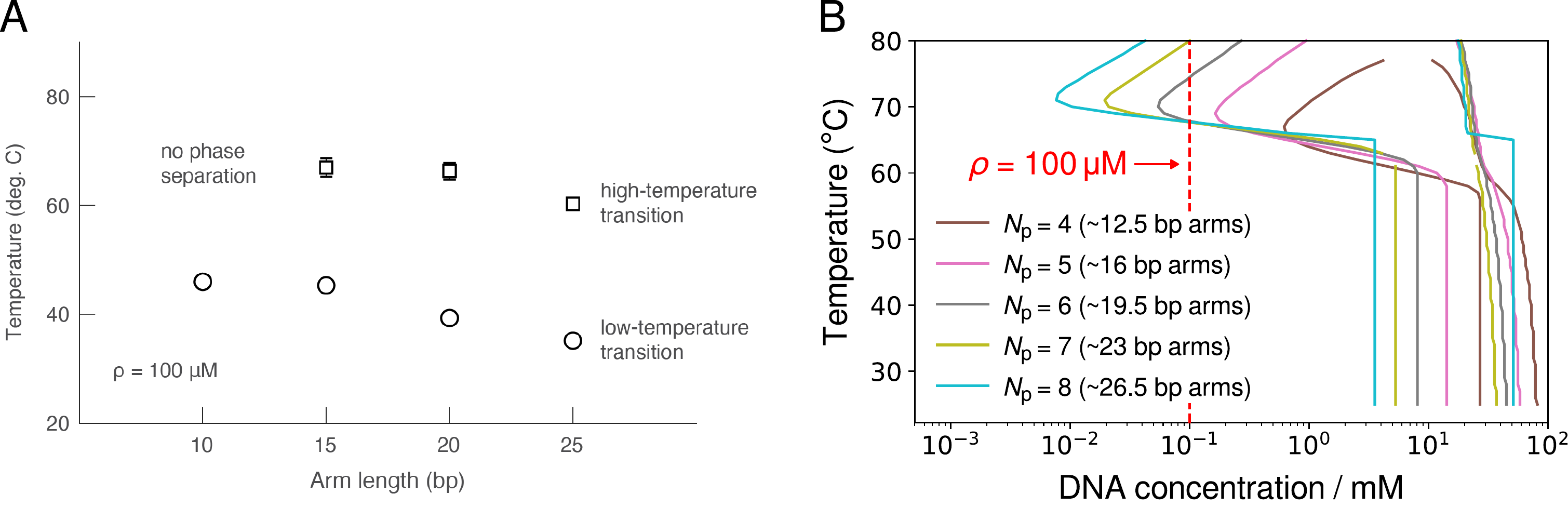}
  \caption{(a)~Experimental measurements of the low- and high-temperature transition temperatures $T_\textrm{trs}$ for stars with different arm lengths. The total DNA concentration is $100\,\mu\text{M}$ and the magnesium concentration is $100\,\text{mM}$. There is no observable high-temperature phase transition for arm lengths of 10 bp.
  (b)~Using the parameterization from the main text, we predict that a high-temperature phase transition will not occur at a total DNA concentration of $100\,\mu\text{M}$ if the arm length is less than $\sim16$ bp ($N_{\text{p}} = 5$).}
  \label{fig:length}
\end{figure}

We characterize the role of the DNA strand length by performing bulk fluorescence microscopy experiments  on DNA nanostars with different arm lengths. We keep the DNA sticky end sequence, the total DNA concentration ($\rho=100\,\mu$M), and the magnesium chloride concentration ([MgCl$_2$=100~mM] constant for all experiments. We vary the arm length from 10~bp to 25~bp in 5-bp steps and characterize the low and high transition temperature $T_\textrm{trs}$ for each star design (\figref{fig:length}a). We find that both the low and high transition temperatures have a very weak dependence on the arm length. The low-temperature transition shifts by roughly $10^\circ$C and the high-temperature transition shifts by at most $7^\circ$C upon increasing the arm length from 15~bp to 25~bp. We highlight that there is no observable high-temperature phase transition for 10-bp-long arms, suggesting that the DNA molecules need to be of a certain length before the disassembled strands can condense at high temperatures.

Interestingly, the predictions of our model, using the parameterization from the main text, are very similar to the experimental observations (\figref{fig:length}b). To account for changes in ssDNA strand length, we vary the number of blobs representing a DNA strand, $N_{\text{p}}$.  We also modify the excluded volume parameter for an assembled nanostar, $N_{\text{s}}$, accordingly, such that $N_{\text{s}} \propto N_{\text{p}}^3$.  Assuming that the nanostar stability depends linearly on the arm length, we then vary the free energy of nanostar formation, $\Delta G$, proportionally to $N_{\text{p}}$.  In this way, we find that the high transition temperature only exists at a total DNA concentration of $100\,\mu\text{M}$ when $N_{\text{p}}$ exceeds 5, which corresponds to an arm length of $\sim$16 bp. Beyond this length, the high transition temperature under experimental conditions is relatively insensitive to the arm length: $T_{\text{trs}}$ varies within 2 degrees when varying the length by a factor of two.

\bibliography{references.bib}